\documentclass[journal]{IEEEtran}
\ifCLASSINFOpdf
\else
\fi
\usepackage[dvipsnames]{xcolor}
\usepackage[normalem]{ulem}
\usepackage{color,soul}
\usepackage{enumitem}
\usepackage{comment}
\usepackage{longtable}
\usepackage{graphicx}
\usepackage[font=normalsize]{caption}
\usepackage{chemformula}
\usepackage{caption}
\usepackage{booktabs}
\DeclareCaptionLabelSeparator{tableNewline}{\par}
\usepackage{caption}
\usepackage{subcaption}
\usepackage[font={small},skip=10pt]{caption}
\usepackage{booktabs}
\usepackage[english]{babel}
\usepackage{multicol}
\usepackage{amsmath}
\usepackage{fancyhdr}
\usepackage[section]{placeins}
\usepackage{multirow}
\usepackage{flushend}
\usepackage{tabularx}
\usepackage{adjustbox}
\usepackage{float}
\usepackage{makecell}
\usepackage{booktabs}
\usepackage{array}
\usepackage{hyperref}
\usepackage{tikz}
\usepackage{quantikz}
\usepackage{amsmath, amssymb} 
\hypersetup{
  colorlinks=true,
  allcolors=blue,
  allbordercolors={blue!50!black}}
\usepackage[noadjust]{cite}
\usepackage[english]{babel}
\usepackage{graphicx}
\usepackage{orcidlink}  
\newcommand{\bluecite}[1]{\textcolor{blue}{\cite{#1}}}

\begin{document}
\bstctlcite{IEEEexample:BSTcontrol}
\setlength{\parskip}{0pt}
\title{Modeling and Analysis of VOC-based Interplant Molecular Communication Channel}
%\title{VOC/Gas Molecule Based Interplant Molecular Communication Channel Model}
\author{Bitop Maitra\scalebox{1.5}{\orcidlink{0000-0003-3948-229X}},~\IEEEmembership{Graduate Student Member,~IEEE}
        and~Ozgur~B.~Akan\scalebox{1.5}{\orcidlink{0000-0003-2523-3858}},~\IEEEmembership{Fellow,~IEEE}
\thanks{The authors are with the Center for neXt-generation Communications (CXC), Department of Electrical and Electronics Engineering, Ko\c{c} University, Istanbul 34450, Turkey (e-mail: \{bmaitra23, akan\}@ku.edu.tr).}
\thanks{O. B. Akan is also with the Internet of Everything (IoE) Group, Electrical Engineering Division, Department of Engineering, University of Cambridge, Cambridge CB3 0FA, UK (email: oba21@cam.ac.uk).}
       
\thanks{This work was supported by the AXA Research Fund (AXA Chair for Internet of Everything at Ko\c{c} University).}
\vspace{-1.2mm}
}

\maketitle

\begin{abstract}
Molecular communication (MC) enables information transfer using particles inspired by biological systems.
Volatile Organic Compounds (VOCs) are one of the most abundant and diverse classes of signaling molecules used by living or non-living objects.
VOC-based MC holds great promise in developing long-range, bio-compatible communication systems capable of interfacing nano- and micro-scale devices.
In this paper, we present a comprehensive end-to-end framework for VOC-based interplant MC from an ICT perspective. The communication process is divided into three stages: transmission (VOC biosynthesis and emission from leaves), channel propagation (advection-diffusion in turbulent wind via Gaussian puff for stress-induced VOC release and Gaussian plume for constitutive VOC release), and reception (VOC uptake and physiological response in the receiver plant). 
Each stage is analyzed by its attenuation and delay. 
Numerical results demonstrate that VOC-based channels exhibit low-pass behavior, with bandwidth and capacity heavily influenced by distance, wind velocity, and noise. 
Though the physical channel supports moderate frequencies, biological constraints at the transmitter restrict the end-to-end channel to slow-varying signals.
\end{abstract}


\renewcommand{\figurename}{Fig.}

\section{Introduction}
\label{sec:Intro}
\IEEEPARstart {M}{olecular} Communication (MC) is an emerging communication paradigm that considers molecular motion as a propagating mechanism, aiming to study biological and chemical communication systems by mimicking them as communication networks and studied through Information and Communication Technology (ICT) \bluecite{farsad2016comprehensive}.
Similar to wireless communication, MC also comprises a transmitter, a channel, and a receiver, where Information Molecules (IMs) are used as the information carrier \bluecite{akyildiz2008nanonetworks}.
However, IMs vary depending on the type of biological or chemical environment; some widely explored IMs are proteins like nucleic acids, ions, chemical messengers, and nanoparticles; whereas, for plants, these IMs are VOCs \bluecite{kuscu2019transmitter}.

\begin{figure}
    \centering
    \includegraphics[width=\linewidth]{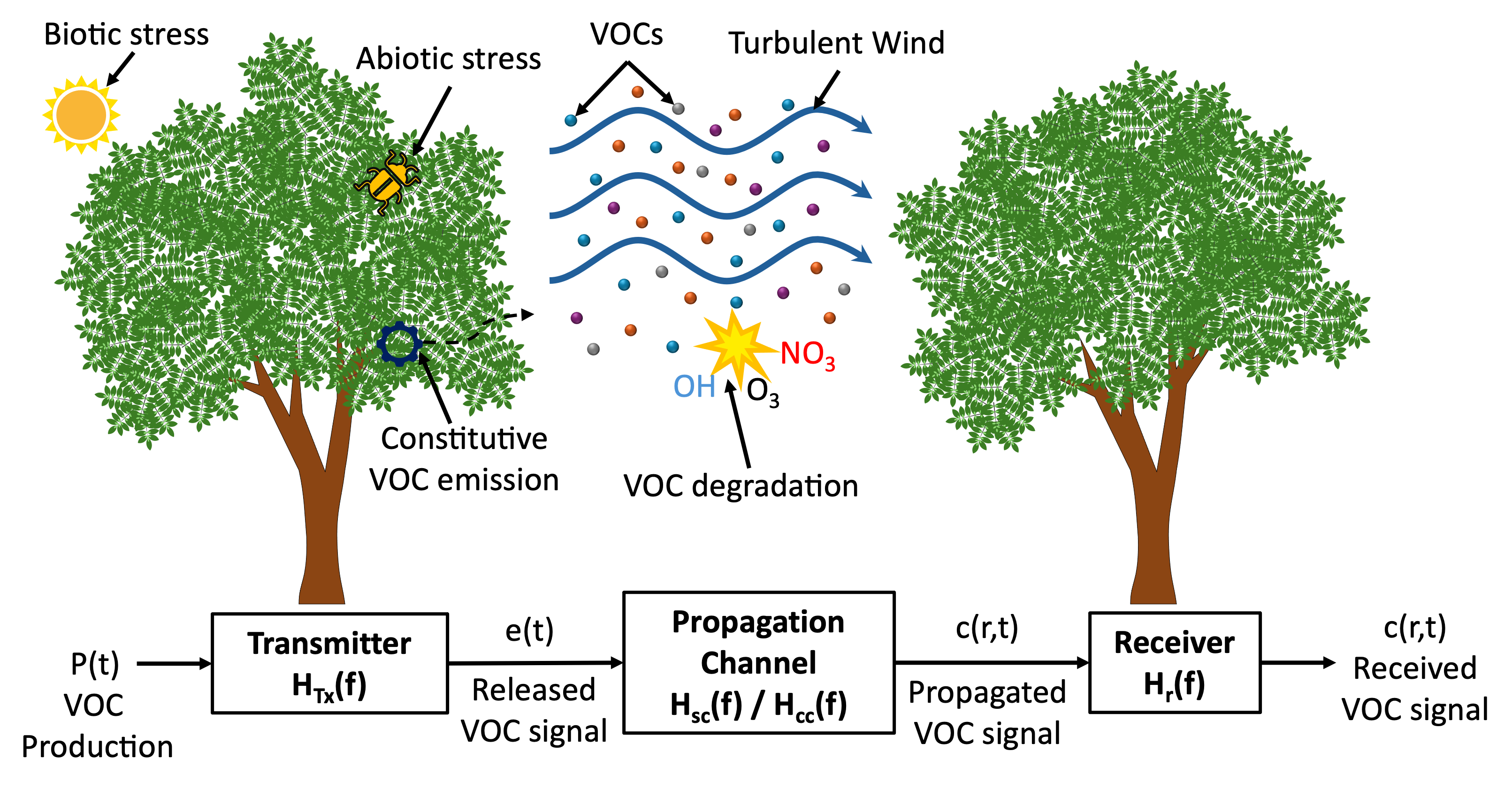}
    \caption{Schematic diagram of VOC-based end-to-end interplant communication.}
    \label{fig:whole_scheme}
\end{figure}

VOCs are low-molecular-weight chemicals that are emitted into the atmosphere by plants as part of their physiological functions.
Plants have developed the ability to produce and emit different VOCs due to the internal or external biotic and abiotic factors that represent the physiological status of the plant \bluecite{dudareva2013biosynthesis}.
%VOCs are used for mating purposes and potentially repelling any predators while attracting mutualists such as beneficial microbes \bluecite{ninkovic2021plant}.
The release of VOCs is of two types, namely constitutive and stress-driven \bluecite{grote2013leaf}.
\begin{comment}
Studies have shown that VOCs play crucial roles in plant-environment and interplant communication under biotic or abiotic stress. 
Stress-driven emissions are often induced and can vary depending on environmental conditions and the nature of the stress \bluecite{dudareva2006plant}.
\end{comment}
It is evident that the constitutively emitted VOCs are released continuously \bluecite{loreto2022plants}, whereas stress-driven VOCs are released almost instantly when abiotic and biotic stress are experienced 
%and act as an immediate alarm signal 
\bluecite{ameye2018green}.
Based on VOCs, neighboring plants communicate, as shown in Fig. \textcolor{blue}{\ref{fig:whole_scheme}}, and the receiver plant can either infer the identity of the emitter plant or VOCs can be used to enhance the resistance systems against harmful interactions with herbivorous insects and environmental conditions, such as drought or mechanical injury \bluecite{ninkovic2016decoding}.

Usually, the impact of individual VOCs like methyl jasmonate or isoprene is studied on receiver plants \bluecite{dong2022exogenous}. 
However, studies show that plants release complex mixtures of VOCs, known as blended VOCs. 
These blends can carry more detailed information than individual VOCs due to the diversity in their composition and relative ratios \bluecite{holopainen2010multiple}. 
%Blended VOCs have been shown to enhance the reliability of the signal received by neighboring plants. 
%For example, receiver plants may respond differently to the same VOC depending on the concentration of different compounds in the blend \bluecite{kigathi2019plant}. 
This emphasizes the importance of studying VOCs as blends.
%Primarily, MC-based communication channels are of limited range, typically in millimeters, and investigate the communication either in nanoscale devices or in cellular environments. However, 
%VOC-based MC communication occurs at the unicellular level, like bacteria \bluecite{wheatley2002consequences}, as well as the multicellular level, like animals \bluecite{fischer2015physiological}. 
VOC-based MC channel exists in the surroundings that can range up to hundreds of meters \bluecite{unluturk2016end}. 
With the potential of long-range communication along with the existence inside all types of living organisms, it paves a way to connect the microscale environment to the macro-world.
%Hence, the application of VOC-based communication is large, such as plant communication, plant bird communication \bluecite{koski2015insectivorous}, stale food or body part and their attractor, pest control, etc. 

\textcolor{black}{Several works have demonstrated advanced VOC-based 
MC systems with experimental validation and pest control applications. 
For instance, field experiments have shown that herbivore-induced VOC emissions enable 
plant-plant and plant-insect signaling, leading to reduced herbivore damage through induced 
defense responses \bluecite{baldwin2006volatile}. 
Similarly, VOCs released by plants under stress 
have been shown to attract natural enemies of herbivores, establishing an indirect pest 
control mechanism via chemical signaling \bluecite{kessler2001defensive}. These studies highlight 
the practical realization of VOC-mediated communication in ecological systems. However, 
such works primarily focus on biological functionality and lack a systematic 
communication-theoretic modeling of realistic propagation, noise, and blended emissions, 
which motivates the proposed framework.
}

In the terrestrial environment, interplant VOC propagation occurs through the ambient air, which acts as the physical communication channel.
VOCs released from plant surfaces diffuse passively under still air conditions, governed by Fick's laws; however, their transmission is dominated by wind-driven advection and turbulent mixing \bluecite{niinemets2013quantitative}. 
Environmental factors like oxidant sources (e.g., hydroxyl radical (OH), nitrate radical (NO\textsubscript{3}), ozone (O\textsubscript{3})) also modulate VOCs' lifetime and transport range \bluecite{waring2015volatile}.
These atmospheric variables influence the concentration, delay, and directionality of VOC transmission. 

VOC-based communication systems from the ICT perspective have not been widely explored. 
However, a simple channel is described without considering the uncertainties or propagation noise in \bluecite{gine2009molecular}. Moreover, an end-to-end VOC propagation channel is modeled in \bluecite{unluturk2016end}. 
This model considers the advection and turbulent diffusion and obtains the channel attenuation and channel delay for the instantaneous release of a single VOC (linalool). 
However, it lacks consideration of noise that is induced by the VOC degradation with the interaction with environmental radicals, and it also doesn't consider the blended VOC release from the transmitter. 
Hence, a comprehensive study is needed to understand how the channel behaves in real environmental conditions for the stress-driven and constitutive release of VOCs.

In this paper, we consider plants as VOC transmitters and receivers, as shown in Fig. \textcolor{blue}{\ref{fig:whole_scheme}}.
%The emission of VOCs may occur via leaves, canopy, and stems; however, the emission from leaves is well documented, and high amounts of VOCs are released from leaves. Hence, 
We consider the leaf-level emission of VOCs from the stomata of a single leaf, which travels through the ambient air media, and the receiver plant uptakes the VOCs through the stomata.
%Upon diffusing into a leaf, the plant activates the physiological responses according to the detected VOCs.
The different steps of VOC communication, i.e., transmission, physical communication channel, and reception, are modeled and analyzed with the normalized gain and delay \textcolor{black}{in the frequency domain}.
\textcolor{black}{The normalized gain quantifies the attenuation of the VOC concentration as they propagate through different media, whereas the delay represents the temporal lag between VOC emission and its detection at the receiver plant.
%These metrics provide a system-level characterization of how efficiently VOC signals propagate between plants in terms of signal strength and propagation time.
}
\textcolor{black}{VOC emissions are dynamic and can be modulated by environmental stresses and circadian rhythms, which induce time-varying VOC emission profiles \bluecite{niinemets2013quantitative, zeng2017regulation}.
Therefore, the frequency domain analysis represents how temporal components of a time-varying emission profile are attenuated and delayed by the transmission, propagation, and reception processes.}

The rest of the paper is as follows.
Sec. \textcolor{blue}{\ref{sec:transmission}} describes the transmission process after the production of VOCs and the release mechanism from the stomata.
In Sec. \textcolor{blue}{\ref{sec:channel}}, the physical channel is modeled considering the stress-driven and constitutive release of VOCs along with a noise model.
Then, the reception process or the uptake process to the receptor plant's leaf is described in Sec. \textcolor{blue}{\ref{sec:reception}}.
Furthermore, Sec. \textcolor{blue}{\ref{sec:analysis}} discusses the numerical results for channel attenuation and delay with physical explanations.
It also briefly describes the end-to-end VOC channel model.
Finally, Sec. \textcolor{blue}{\ref{sec:conclusion}} concludes this paper with a future direction of VOC-based MC.

\vspace{-0.3cm}
\section{VOC Production and Transmission Model}
\label{sec:transmission}
\begin{comment}
\begin{align}
  s(t) = s_c(t) ((1-\beta) + \beta e^{-\pi{t^2}}),
  \label{s(t)}
\end{align}
where $S_c$ is the continuous VOC production rate, 
$\beta(t)$ is the binary switching function to account the continuous and stress driven MC, it is 0 for continuous
VOC release and 1 for stress driven VOC release.
$e^{-\pi{t^2}}$ is the gaussian pulse.
\end{comment}

\begin{figure}[t!]
    \centering
    \includegraphics[width=0.94\linewidth]{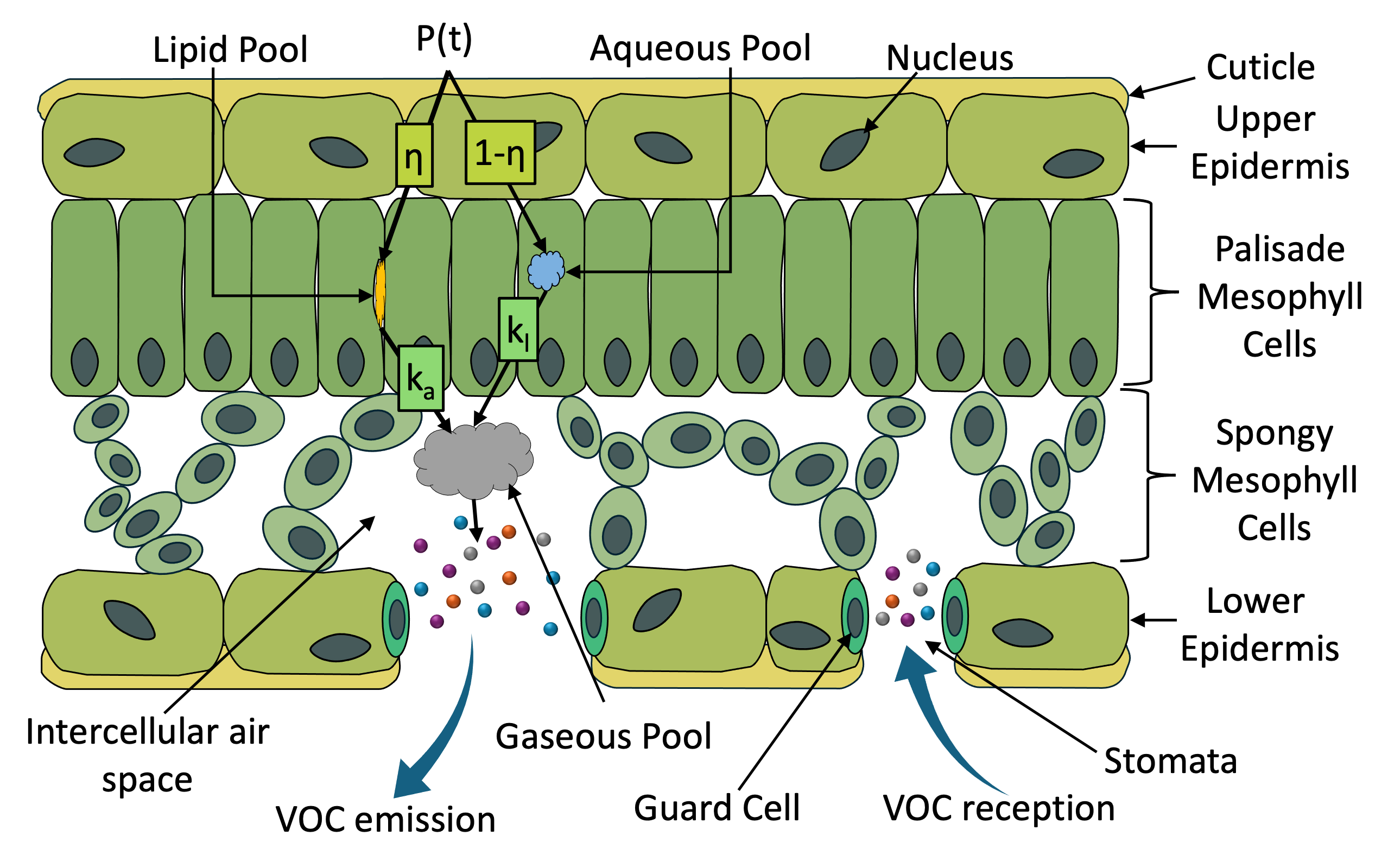}
    \caption{Schematic diagram of leaf cross-section showing VOC production, emission, and reception through stomata \bluecite{harley2013roles}.}
    \label{fig:stomata}
\end{figure}

In this study, we consider that the VOCs are released from a leaf of a plant. 
%A leaf consists of an upper and lower epidermis, each usually one cell thick and covered by a water-proof cuticle made of cutin to minimize water loss. 
%Beneath the upper epidermis lies the palisade mesophyll comprising elongated chloroplast-rich cells. The spongy mesophyll, made up of loosely packed and spherical cells with large intercellular air spaces that facilitate gas exchange, is found underneath. 
Stomata, consisting of a pore guarded by two guard cells, regulate the exchange of VOCs, as shown in Fig. \textcolor{blue}{\ref{fig:stomata}}.
Specialized secretory cells and structures, like trichomes, idioblasts, cavities, and secretory ducts, synthesize VOCs in plants, such as terpenes and essential oils \bluecite{FAHN200037}. 

\begin{comment}
These compounds are synthesized in various subcellular compartments, mainly in plastids, often leucoplasts devoid of ribosomes and thylakoids for monoterpenes, and sometimes in mitochondria and Endoplasmic Reticulum (ER). 
In some species, mitochondria have been directly involved in VOC synthesis. 
Periplastidic ER surrounds the plastids and may function as a transport duct for the synthesized VOCs.
The ER and sometimes Golgi-derived vesicles mediate the intracellular trafficking of VOCs toward the plasma membrane.
Once transported, the VOCs are secreted through two main mechanisms: eccrine secretion, i.e., direct molecular or ionic transport across membranes, and granulocrine secretion, i.e., via fusion of vesicles with the plasma membrane.
Finally, depending on the structure of the secretory tissue, the VOCs accumulate in subcuticular spaces or are released through pores in the cuticle.
The secretory cells are supported by stalk cells with completely cutinized side walls, which prevent the backflow of accumulated VOCs into the plant \bluecite{FAHN200037}.
\end{comment}

After the production of VOCs, i.e., $P(t)$, they are partitioned according to their solubility, as shown in Fig. \textcolor{blue}{\ref{fig:stomata}}. 
If the VOCs are water soluble, they are stored in aqueous pools, \textcolor{black}{where $S_a(t)$ is the amount of VOCs present in the aqueous storage pool.}
The rate of release of the VOCs from the aqueous pools can be described by the first-order kinetics equation as \bluecite{harley2013roles}:
\begin{align}
\frac{dS_a(t)}{dt} = \eta P(t) - k_a S_a(t),
\label{s_a}
\end{align}
where $k_a$ is the aqueous storage pool VOC-specific first-order kinetic constant, and $\eta$ is the partitioning coefficient that depends on the solubility and the diffusion coefficient of the VOCs. 
On the other hand, if VOCs are lipid soluble, they are stored in lipid pools, \textcolor{black}{where $S_l(t)$ denotes the amount of VOCs present in the lipid storage pool.}
The release rate of these lipid soluble VOCs can be expressed as \bluecite{harley2013roles}:
\begin{align}
\frac{dS_l(t)}{dt} = (1 - \eta) P(t) - k_l S_l(t),
\label{s_l}
\end{align}
where $k_l$ is the lipid storage pool VOC-specific first-order kinetic constant.
The VOCs of these pools then diffuse to the intercellular air space of the leaf before emission to the ambient air medium. 
The released VOCs are stored in the gas phase pools, \textcolor{black}{where $S_g(t)$ denotes the amount of VOCs present in the gas storage pool.}
From the gas phase pools, the release rate of VOCs to the ambient air can be expressed as \bluecite{harley2013roles}:
\begin{align}
\frac{dS_g(t)}{dt} = k_a S_a(t) + k_l S_l(t) - k_g S_g(t),
\label{s_g}
\end{align}
where $k_g$ is the gas storage pool kinetic constant.
\textcolor{black}{(\ref{s_a}-\ref{s_g}) describe the internal VOC partitioning and transport dynamics within the transmitter plant prior to the emission into the propagating medium.}
The VOC emission rate ($e(t)$) from the leaf represents the flux of VOCs from the gas phase pool to the ambient air through stomata, and can be expressed as \bluecite{harley2013roles}:
\begin{align}
e(t) = k_g S_g(t),
\label{q}
\end{align}
The Fourier transform of the VOC emission rate \textcolor{black}{with respect to time `$t$'} is given by:
\begin{align}
E(f) = k_g S_g(f),
\label{Qf}
\end{align}
From (\ref{s_a}-\ref{Qf}), we can deduce the transfer function ($H_{Tx} (f)$) \textcolor{black}{by considering VOC emission flux from the transmitter plant as the system output, while the VOC production in the plant leaf represents the system input (please refer to Appendix \ref{A_A} for the derivation)}:
\begin{align}
  H_{Tx} (f) & = \frac{k_g}{j2 \pi f + k_g} \times \left(\frac{k_a \eta}{j2 \pi f + k_a} + \frac{k_l (1 - \eta)}{j2 \pi f + k_l} \right),
\end{align}
%The transfer function, $H_{Tx}(f)$, depends on the solubility coefficient $\eta$. 
%However, we are interested in determining the normalized gain as it shows how the signal is attenuated during the diffusion from the intercellular air space to the ambient air. 
The normalized gain ($H_{Tx}^{norm}(f)$) of the transmission can be \textcolor{black}{defined as the magnitude of the transfer function divided by its zero-frequency magnitude}, expressed as:
\begin{comment}
\begin{align}
& H_{Tx}^n(f)  = \frac{|H_{Tx}(f)|}{max |H_{Tx} (f)|} \nonumber \\
&= \frac{k_g}{\sqrt{k_g^2 + (2\pi f)^2}} \left ( \frac{k_a \eta}{\sqrt{k_a^2 + (2\pi f)^2}} + \frac{k_l (1-\eta)}{\sqrt{k_l^2 + (2\pi f)^2}}\right)
\label{NGT}
\end{align}
\end{comment}
\begin{align}
H_{Tx}^{norm}(f) = \frac{k_g}{\sqrt{k_g^2 + (2\pi f)^2}} \sqrt{\mathrm{Im}^2 (f) + \mathrm{Re}^2 (f)},
\label{NGT}
\end{align}
where 
\begin{align}
    \mathrm{Im} (f) = \left ( \frac{k_a \eta \cdot 2 \pi f}{k_a^2 + (2 \pi f)^2} + \frac{k_l (1- \eta) \cdot 2 \pi f}{k_l^2 + (2 \pi f)^2} \right ), \\ 
    \mathrm{Re}(f) = \left(\frac{k_a^2 \eta}{k_a^2 + (2 \pi f)^2} + \frac{k_l^2 (1-\eta)}{k_l^2 + (2 \pi f)^2} \right),
\end{align}
Moreover, the phase of $H_{Tx}(f)$ is expressed as:
\begin{comment}
\begin{align}
    \phi_{Tx} (f)  & = tan^{-1} \left(\frac{Im(H_{Tx}(f))}{Re(H_{Tx}(f))}\right), \nonumber & \\ 
     & = tan^{-1} \left(-\frac{2 \pi f}{k_g}\right) + tan^{-1} \left(-\frac{2 \pi f}{k_a}\right) \nonumber & \\ & \quad \quad \quad \quad \quad \quad \quad \quad \quad + tan^{-1} \left(-\frac{2 \pi f}{k_l}\right),
\end{align}
\end{comment}
\begin{align}
    \phi_{Tx} (f)  & = tan^{-1} \left(\frac{\mathrm{Im}(H_{Tx}(f))}{\mathrm{Re}(H_{Tx}(f))}\right), \nonumber & \\ 
     & = tan^{-1} \left(-\frac{2 \pi f}{k_g}\right) + tan^{-1} \left(- \frac{\mathrm{Im}(f)}{\mathrm{Re}(f)}\right),
\end{align}
Hence, the delay can be obtained as:
\begin{comment}
\begin{align}
    \tau_{Tx} (f) = -\frac{d \phi_{Tx} (f)}{df} = \frac{2\pi k_g}{k_g^2 + (2\pi f)^2} + \frac{2\pi k_a}{k_a^2 + (2\pi f)^2} \nonumber & \\ + \frac{2\pi k_l}{k_l^2 + (2\pi f)^2},
\end{align}
\end{comment}
\begin{align}
    \tau_{Tx} (f) = & -\frac{d \phi_{Tx} (f)}{df} = \frac{2\pi k_g}{k_g^2 + (2\pi f)^2} + \nonumber  \\ & \frac{\mathrm{Re}(f) \cdot \mathrm{Im}'(f) - \mathrm{Im}(f) \cdot \mathrm{Re}'(f)}{\mathrm{Re}^2(f) + \mathrm{Im}^2 (f)},
\end{align}
where $\mathrm{Re}'(f)$ and $\mathrm{Im}'(f)$ are the first order derivative of $\mathrm{Re}(f)$ and $\mathrm{Im}(f)$. 
\begin{comment}
, and given by:
\begin{align}
    \mathrm{Re}'(f) = -\frac{8 \pi^2 k_a^2 \eta f}{(k_a^2+(2 \pi f)^2)^2} - \frac{8 \pi^2 k_l^2 (1-\eta) f}{(k_l^2 + (2 \pi f)^2)^2},
\end{align}
\begin{align}
    \mathrm{Im}'(f) &= \frac{2\pi k_a \eta (k_a^2-(2 \pi f)^2) }{(k_a^2 + (2 \pi f)^2)^2} + \nonumber \\ &\hspace{6.5em} \frac{2 \pi (1-\eta) k_l (k_l^2 - (2 \pi f)^2)}{(k_l^2 + (2 \pi f)^2)^2},
\end{align}
\end{comment}

\section{VOC Propagation and Channel Modeling}
\label{sec:channel}
After the release of VOCs from the leaf, they start to propagate in the medium.
Depending upon the environment, the medium can be water for aquatic plants and air for terrestrial plants.
However, in this section, we will consider air as the propagating medium and describe the VOC propagation depending on the VOC emission type.
 
To begin with, we initialize the modeling with some assumptions. 
We consider a constant wind velocity ($\vec{u}$) in the medium.
The wind flow is classified according to Pasquill-Gifford (PG) into 6 stability classes: A (strongly unstable), B (moderately unstable), C (slightly unstable), D (neutral), E (slightly stable), and F (moderately stable) \bluecite{turner1994atmospheric}.
\textcolor{black}{However, the D stability class is typically observed under moderate wind and cloudy conditions. 
It is commonly used as a reference condition in atmospheric dispersion modeling \bluecite{turner1994atmospheric}.}
Hence, we consider class D as our wind stability, whose velocity typically varies from 3-7 $m/s$ \bluecite{turner1994atmospheric}. 
\begin{comment}
Furthermore, we consider the channel to be a time-invariant model to ensure predictability and stability over time.
The model will be developed considering a maximum distance of 200 meters between the plants, and the least velocity of the model will be 3 $m/s$, as stated earlier.
Hence, the time delay ($\tau_d$) will be $\tau_d = 500/3 =166.667 s$.
In order to make our assumption of time invariance valid, we need to have the time step greater than the time delay.
One widely accepted time step is 10 min \bluecite{turner1994atmospheric}, which is greater than the time delay.
Hence, the channel model can be considered as time-invariant.
\end{comment}

%The VOCs emitted from the leaf start to propagate through the medium.
The propagation of VOCs is governed by Fick's law of diffusion, which is well documented in the literature \bluecite{gine2009molecular}, given by:
\begin{align}
    \frac{\partial c(\vec{r}, t)}{\partial t} + \nabla \cdot \vec{J}(\vec{r}, t) = S (\vec{r},t),
    \label{Fick}
\end{align}
where $c(\vec{r},t)$  is the concentration at a position $\vec{r}  \in \mathbb{R}^3$, $\nabla$ is the Laplace operator in 3-D Cartesian coordinates, $\vec{J}(\vec{r}, t)$ is the flux of VOCs at position $\vec{r}$, and $S (\vec{r},t)$ is the source term.

The flux of VOCs ($\vec{J}(\vec{r}, t)$) depends on both the advection due to the wind and diffusion caused by the turbulence in the atmosphere, expressed as:
\begin{align}
    \vec{J}(\vec{r}, t) = \vec{J}_A(\vec{r}, t) + \vec{J}_D(\vec{r}, t),
    \label{Flux}
\end{align}
\textcolor{black}{In this study, the wind velocity is considered to be a constant ($\vec{u}$), and the effects of atmospheric turbulence are incorporated via eddy diffusion that governs the dispersions in the advection-diffusion equation.}
Hence, the advection term can be expressed as: 
\begin{align}
    \vec{J}_A(\vec{r}, t) = c (\vec{r}, t) \vec{u},
    \label{advection}
\end{align}
Furthermore, the flux due to the eddy diffusion ($K$) is expressed as:
\begin{align}
    \vec{J}_D(\vec{r}, t) = -K \nabla c(\vec{r},t),
    \label{Eddy}
\end{align}
By substituting (\ref{Flux}-\ref{Eddy}) in (\ref{Fick}), the obtained expression is:
\begin{align}
    \frac{\partial c(\vec{r}, t)}{\partial t} + \nabla \cdot (c (\vec{r}, t) \vec{u}) - \nabla \cdot (K \nabla c(\vec{r},t)) = S (\vec{r},t), 
    \label{final}
\end{align}
(\ref{final}) needs to be solved in order to obtain the closed-form solution considering suitable initial and boundary conditions. 
%However, it is important to note that the solution of the PDE may not always exist.
\textcolor{black}{Since (\ref{final}) corresponds to a linear advection-diffusion equation with constant wind velocity and strictly positive diffusion coefficients, it is uniformly parabolic and therefore results in a unique solution under appropriate initial and boundary conditions \bluecite{evans2022partial}.
The fundamental solution of such a linear equations is Gaussian, which is the basis of classical atmospheric dispersion models \bluecite{stockie2011mathematics}.
}
For the stress-driven VOC release, we consider the Gaussian Puff model, as it deals with the sudden release of a fixed amount of VOCs.

%In order to solve the PDE, we will consider some assumptions based on the type of VOC emission.

%After the release, the propagation of the VOCs depends on the medium and also the type of release.
%According to the release type, the propagation can be modeled via Gaussian Models. 
%The use of Gaussian models in air-based propagation is well documented \bluecite{lotrecchiano2020pollution, arystanbekova2004application}.

\begin{comment}
\subsection{Channel Modeling for Constitutive VOC Emission}
The plume model is governed by \bluecite{snoun2023comprehensive, horst1977surface}:
\begin{align}
    c(\vec{r}, t) = \frac{Q_c}{2 \pi u \sigma_y \sigma_z}  e^{-\frac{1}{2}\left( \frac{y-y_0}{\sigma_y} \right)^2} 
      \left[ e^{- \frac{{(z-z_0)}^2} {2\sigma_z^2}} + e^{- \frac{(z+z_0)^2}{2\sigma_z^2}}\right]
      \label{puff_ori}
\end{align}

The fourier transform is given by:
\begin{align}
    H_{cc}(f)= \frac{Q_c}{2 \pi u \sigma_y \sigma_z} & e^{-\frac{1}{2}\left( \frac{y-y_0}{\sigma_y} \right)^2} \nonumber \\
     & \left[ e^{- \frac{{(z-z_0)}^2} {2\sigma_z^2}} + e^{- \frac{(z+z_0)^2}{2\sigma_z^2}}\right] \delta(f)
\end{align}

The normalised gain can be expressed as:
\begin{align}
    H^n_{cc}(f) = \delta(f)
\end{align}

Moreover, the phase is:
\begin{align}
    \phi_{cc} (f) = 0
\end{align}
Following $\phi_{cc}(f)$, the phase delay is 0.
\end{comment}

\subsection{Channel Modeling for Stress-driven VOC Emission}
\label{subsec:puff}
We consider some assumptions in order to model the puff concentration profile, where the wind velocity is constant and also directed in the $x$-axis ($u,0,0$).
According to Taylor's theory, the dispersion coefficients are used to describe the effects of eddy diffusion in atmospheric transport modeling \bluecite{taylor1922diffusion}.
In realistic atmospheric conditions, the diffusion is anisotropic and depends upon the downwind distance $x$.
The governing equation for the puff model is expressed as \bluecite{li2019computational}, (\textcolor{black}{please refer to Appendix \ref{A_B} for the derivation}):
\begin{align}
    c(\vec{r}, t) = \frac{Q_0}{(2 \pi)^\frac{3}{2} \sigma_x \sigma_y \sigma_z} e^{-\frac{1}{2}  \left( \frac{x-x_0-ut}{\sigma_x} \right)^2}  e^{-\frac{1}{2}\left( \frac{y-y_0}{\sigma_y} \right)^2} \nonumber \\
      \left[ e^{- \frac{{(z-z_0)}^2} {2\sigma_z^2}} + e^{- \frac{(z+z_0)^2}{2\sigma_z^2}}\right],
      \label{puff_ori}
\end{align}
where $c(\vec{r}, t)$ is the spatiotemporal concentration, $Q_0$ is the number of released VOCs, and $\sigma_x$, $\sigma_y$, and $\sigma_z$ are the dispersion coefficients in the $x$, $y$, and $z$ direction, respectively. $x_0$, $y_0$, $z_0$ are the 3D coordinate of the VOC release point.
Furthermore, we consider that VOCs are released from a certain height $z_0$, i.e., ($0,0,z_0$).
The longitudinal diffusion is negligible compared to advection and is often omitted in the Gaussian puff models \bluecite{unluturk2016end}.
However, this does not imply the absence of dispersion along the x-direction. 
In reality, turbulent eddies induce a finite and small longitudinal spread. 
To capture this behavior, we assume the longitudinal dispersion as the geometric mean of lateral and vertical spreads \textcolor{black}{as they jointly influence the puff development \bluecite{turner1994atmospheric}, expressed as: $\sigma_x = \sqrt{\sigma_y \sigma_z}$.}
%\textcolor{blue}{This assumption is consistent with the shear dispersion theory, 
%where longitudinal spreading arises from the interaction of velocity gradients and diffusion. }
Hence, (\ref{puff_ori}) reduces to:
\begin{align}
    c(\vec{r}, t) = \frac{Q_0}{(2 \pi\sigma_y \sigma_z)^\frac{3}{2}} e^{-\frac{1}{2}  \left( \frac{x-ut}{\sqrt{\sigma_y \sigma_z}} \right)^2}  e^{-\frac{1}{2}\left( \frac{y}{\sigma_y} \right)^2} \nonumber \\
      \left[ e^{- \frac{{(z-z_0)}^2} {2\sigma_z^2}} + e^{- \frac{(z+z_0)^2}{2\sigma_z^2}}\right],
      \label{puff_reduced}
\end{align}
The dispersion coefficient depends on the turbulent airflow and the stability class. 
There are multiple ways to determine these coefficients, like Briggs \bluecite{briggs1973diffusion}, Briggs dispersion coefficients that are revised by Griffiths \bluecite{griffiths1994errors}, and computational methods like genetic algorithm and particle swarm optimization \bluecite{abualigah2015applying}. For the D stability class, $\sigma_y$ and $\sigma_z$ are $0.06x^{0.92}$ and $0.15x^{0.70}$, respectively \bluecite{griffiths1994errors}.
\begin{comment}
\begin{table}[t!]
\centering
\caption{Puff dispersion coefficient formulas \bluecite{li2019computational}}
\begin{tabular}{|c|c|c|}
\hline
\textbf{PG stability class} & \(\boldsymbol{ \sigma_y \, (\text{m})}\) & \(\boldsymbol{\sigma_z \, (\text{m})}\) \\ 
\hline
A & \(0.18x^{0.92}\) & \(0.60x^{0.75}\) \\ 
B & \(0.14x^{0.92}\) & \(0.53x^{0.73}\) \\ 
C & \(0.10x^{0.92}\) & \(0.34x^{0.71}\) \\ 
D & \(0.06x^{0.92}\) & \(0.15x^{0.70}\) \\ 
E & \(0.04x^{0.92}\) & \(0.10x^{0.65}\) \\ 
F & \(0.02x^{0.89}\) & \(0.05x^{0.61}\) \\ \hline
\end{tabular}
\label{tab:pg_dispersion}
\end{table}
\end{comment}
However, during the propagation, VOCs are influenced by two modalities: chemical interference and air turbulence \bluecite{wilson2015noisy}. 
The Gaussian models inherently consider the air turbulence via the dispersion coefficients \bluecite{cao2011dispersion}.
Hence, we need to consider the chemical interference of VOCs.

Initially, depending on the type of VOCs, they react with OH radicals,  NO\textsubscript{3} radicals, and O\textsubscript{3} atoms or undergo photodissociation in the troposphere and form alkyl or substituted alkyl radicals ($R^a$) (e.g., hydroxyalkyl, nitrooxyalkyl, or oxoalkyl).
Furthermore, those $R^a$s undergo a sequential reaction to produce a stable product \bluecite{atkinson2003atmospheric}. 
The equation can be expressed as: $\mathrm{VOC} + (\mathrm{OH}/\mathrm{NO_3}/\mathrm{O_3}/h\nu) \to \mathrm{R^a}$.
As soon as $R^a$s are produced in the troposphere, the properties of VOCs are lost, and those are no longer considered as the information carrier in the interplant communication process.
\textcolor{black}{Reaction intermediates are not explicitly modeled as separate species; their formation is treated as irreversible removal of the parent VOC from the information carrying pool.} 
%Since byproducts do not participate in interplant signaling, subsequent reactions are beyond the scope of the communication channel model.
The rate of the reaction ($R$) can be given by: $R = k[VOC][oxidant]$,
where $k$ is the second-order rate constant, and $[VOC]$ and $[oxidant]$ are the concentrations of VOC and oxidants (OH/NO\textsubscript{3}/O\textsubscript{3}), respectively.
Considering the VOC concentration at $\vec{r}$, the rate of the reaction becomes:
\begin{align}
    R = k_{eff}c(\vec{r},t),
    \label{chem_deg}
\end{align}
where $k_{eff} = \sum_i k_i{[oxidant]}_i$ \bluecite{atkinson2003atmospheric}. 
\textcolor{black}{This holds under conditions where the oxidants act independently, and their concentration varies slowly relative to the reaction timescale.}
Considering this VOC degradation, an \textcolor{black}{ordinary} differential equation (ODE) can be expressed as:
\begin{align}
    \frac{d c}{dt} = -k_{eff}t,
    \label{deg}
\end{align}
Considering the solution of (\textcolor{blue}{\ref{deg}}), the concentration profile is given by:
%\begin{align}
 %   c_d(\vec{r},t) = c(\vec{r},t) e^{-k_{eff}t},
%\end{align}
%Hence, the concentration profile considering the VOC degradation is expressed as:
\begin{align}
    c_d(\vec{r}, t) = \frac{Q_0}{(2 \pi\sigma_y \sigma_z)^\frac{3}{2}} e^{-\frac{1}{2}  \left( \frac{x-ut}{\sqrt{\sigma_y \sigma_z}} \right)^2}  e^{-\frac{1}{2}\left( \frac{y}{\sigma_y} \right)^2} e^{-k_{eff}t}\nonumber \\
      \left[ e^{- \frac{{(z-z_0)}^2} {2\sigma_z^2}} + e^{- \frac{(z+z_0)^2}{2\sigma_z^2}}\right],
      \label{VOC_deg_conc}
\end{align}
\textcolor{black}{The transfer function of stress-driven VOC propagation channel ($H_{sc}(f)$) is obtained with an impulse as the input and the Fourier transform of $c_d(\vec{r}, t)$ with respect to time `$t$' as the output, is given by:}
\begin{align}
    H_{sc}(f) = \frac{Q_0}{2\pi u \sigma_y \sigma_z} e^{-\frac{y^2}{2 \sigma_y^2}}  \left[e^{\frac{-(z-z_0)^2}{2 \sigma_z^2}} + e^{\frac{-(z+z_0)^2}{2 \sigma_z^2}} \right]  \nonumber \\
    e^{(k_{eff}-i2 \pi f )\frac{x}{u}} \cdot e^{(k_{eff}- i 2 \pi f)^2\frac{\sigma_y \sigma_z}{u^2}},
\end{align}
\textcolor{black}{Gain ($H^g_{sc} (f)$) can be defined as the magnitude of $H_{sc}(f)$:}
\begin{align}
    H^g_{sc} (f) = \frac{Q_0}{2\pi u \sigma_y \sigma_z} e^{-\frac{y^2}{2 \sigma_y^2}}  \left[e^{\frac{-(z-z_0)^2}{2 \sigma_z^2}} + e^{\frac{-(z+z_0)^2}{2 \sigma_z^2}} \right] \nonumber \\
     \cdot e^{k_{eff} \frac{x}{u} + (k_{eff}^2 - (2\pi f)^2)\frac{\sigma_y \sigma_z}{u^2}},
\end{align}
The normalized gain ($H^{norm}_{sc} (f)$) is given by:
\begin{align}
    H^{norm}_{sc} (f) = \frac{|H^g_{sc} (f)|} {max_f |H^g_{sc} (f)|},
    \label{norm_gain}
\end{align}
\begin{comment}
The phase can be expressed as:
\begin{align}
    \phi_{sc} (f) = -2 \pi f \frac{x}{u} - 4 \pi f k_{eff} \frac{\sigma_y \sigma_z}{u^2},
\end{align}
\end{comment}
Also, the delay will be:
\begin{align}
    \tau_{sc} (f) = -\frac{\phi_{sc} (f)}{df} =2 \pi \frac{x}{u} + 4 \pi k_{eff} \frac{\sigma_y \sigma_z}{u^2},
\end{align}

\subsection{Channel Modeling for Constitutive VOC Emission}
\label{subsec:plume}
The constitutive VOC emission is evident to be released continuously \bluecite{loreto2022plants}, and the continuous VOC emission can be modeled by the Gaussian plume model.
The Gaussian plume is considered to be a borderline case of puffs and the combination of multiple puffs that are close to each other \bluecite{snoun2023comprehensive}.
Hence, the concentration profile of the plume can be expressed as $c_p(\vec{r}) = \int_0^\infty c_d(\vec{r},t) dt$, where $c_d(\vec{r},t)$ is modeled in (\textcolor{blue}{\ref{VOC_deg_conc}}). 
By solving, the obtained concentration profile is:
\begin{align}
    c_p(\vec{r}) = \frac{Q_0}{2 \pi u \sigma_y \sigma_z}   e^{- \frac{y^2}{2 \sigma_y^2} } 
      \left[ e^{- \frac{{(z-z_0)}^2} {2\sigma_z^2}} + e^{- \frac{(z+z_0)^2}{2\sigma_z^2}}\right] e^{-k_{eff}\frac{x}{u}},
      \label{plume}
\end{align}
\textcolor{black}{For the continuous VOC emission, the plume concentration $c_p(\vec{r})$ corresponds to the steady-state response of the previous stress-driven system.
Hence, the transfer function is given by:} $H_{ccs}(f) = c_p(\vec{r}) \cdot \delta(f)$,
Here, the Fourier transform is only defined at $f=0$. 
Even though the continuous VOC release is considered, the amount of VOC changes due to environmental factors like circadian rhythm and weekly or seasonal variation.  
Hence, we intend to model the channel to capture this effect by considering VOC variation in a 24-hour window. Hence, $\Delta t = 86400$s.
According to the time-frequency uncertainty principle $\Delta t \cdot \Delta f \geq \frac{1}{4\pi}$ \bluecite{oppenheim2013human}.
Hence, $\Delta f \geq 9.21 \times 10^{-6}$ Hz.
With this consideration, $H_{ccs}(f)$ becomes:
\begin{align}
    H_{cc}(f) = c_p(\vec{r}) \cdot G(f),
    \label{modified fourier plume}
\end{align}
where $G(f)$ is the Gaussian approximated delta function, i.e., $G(f) = \frac{1}{\sqrt{2 \pi} \Delta f} e^{-\frac{f^2}{2 \Delta f^2}}$
%\begin{align}
 %   G(f) = \frac{1}{\sqrt{2 \pi} \Delta f} e^{-\frac{f^2}{2 \Delta f^2}},
%\end{align}
The normalized gain ($H_{cc}^{norm}(f)$) is given by:
\begin{align}
    H_{cc}^{norm}(f)  & = \frac{|H_{cc}(f)|}{max |H_{cc} (f)|},
\end{align}
\begin{comment}
\textcolor{red}{The delay for this case is $0$, as (\textcolor{blue}{\ref{modified fourier plume}}) doesn't have any phase.
This is explained as, for the continuous VOC emission, VOCs are always present in the medium, and there is no distinct/instantaneous emission like stress-driven VOC emission.}
\textcolor{blue}{Although VOC transport involves finite physical propagation time, the continuous emission establishes a steady-state concentration field. As a result, any variation in emission is effectively reflected immediately in the received signal. Since the channel’s frequency response is real and lacks a phase component, the group delay is mathematically zero, which aligns with the steady-state assumption.
Although VOC transport requires finite time to reach the receiver, continuous emission during the daytime can lead to a quasi-steady-state concentration field. However, due to circadian rhythms and environmental modulation, the emission rate — and thus the concentration field — exhibits diurnal variation, with higher levels during the day and lower levels at night.}
\end{comment}
(\textcolor{blue}{\ref{modified fourier plume}}) does not explicitly contain any phase term and, therefore, doesn't produce any delay.
This is because 
%the plume model considers that VOCs have been emitted continuously for an infinite time, and 
the system is in a quasi-steady state at the time of observation. 
%However, the absence of the delay term doesn't imply instantaneous transport of VOCs.
%It can be explained as the receiver is continuously exposed to a VOC concentration that integrates the cumulative effects of past emissions. 
It is important to note that even though the continuous emission doesn't yield a delay, the delay is implicitly encoded in the derivation from the time-dependent Gaussian puff model, which is explained in Sec. \textcolor{blue}{\ref{subsec:puff}}.
For this reason, the delay is considered to be zero for continuous emission, not due to the physical absence of propagation time but due to the quasi-steady nature of the model.

%However, VOC transpo

\subsection{Noise Model}
\label{noise}
%In ICT, noise is any signal other than the signal of interest that influences a receiver.
Like other communication techniques, VOC communication is also prone to noise, especially when we discuss long-range VOC communication.
We usually discuss this on the basis of a single plant being used as a transmitter and another plant as a receiver.
However, in the environment, multiple plant species are available, and their response to biotic/abiotic stresses widely differ \bluecite{kesselmeier1997emission}. 
Hence, while interplant communication occurs within two plants, other nearby plants \textcolor{black}{of other species}, which are exposed to similar or any other kind of stress, would also release VOCs as a response.
\textcolor{black}{The VOCs, released by the plants of other species, may be considered as noise to the native transmitter-receiver pair.}
\begin{comment}
\textcolor{red}{The constitutive VOC emission is continuous, where VOC concentration is always available in the channel and propagates toward the receiver. 
In this case, as there is no delay, the receiver plant can continuously receive the VOCs, and can approximately infer regarding the transmitter plant's information even in the noisy environment.}
\end{comment}
%Under constitutive emission, VOCs are released continuously by the transmitter plant, which results in a continuous circadian rhythm following a concentration profile \bluecite{zeng2017regulation}. 
%While there is a finite delay due to transport mechanisms, the receiver plant may continuously sense VOC presence and potentially infer transmitter state or identity information, even in noisy environments.
%However, as stress-driven VOC release is instantaneous, it is highly influenced by the VOCs of different plant species, which is not the point of interest in the model; hence, it is considered as noise. 
\textcolor{black}{Even though VOC emission is inherently broadcast into the medium, advection introduces a directional bias in propagation, resulting in stronger transport in the wind direction.
Hence, the communication is analyzed using a transmitter-receiver pair aligned with the dominant flow.}

The trade-off between the signal and the noise depends on the induced stress, the type of plant species, the ratio/number of the sources, and the distance among them. 
The signal-to-noise ratio (SNR) can be expressed as \bluecite{kilinc2014receiver}:
\begin{align}
    SNR = \frac{P_0}{P_N},
\end{align}
where $P_0$ is the average power of the signal, and $P_N$ is the average power of the noise.
Considering the same height of the noise release point, SNR becomes:
\begin{comment}
\begin{align}
    SNR = \frac{Q_0}{Q_N} \cdot e^{ \left ( \frac{x_0^2}{2 \sigma_y \sigma_z} - \frac{x-ut}{{\sigma_y \sigma_z}} x_0 \right )} e^{\left ( \frac{y_0^2}{2 \sigma_y^2} - \frac{yy_0}{\sigma^2} \right )} e^{-(k_{eff}-k_N)t}
\end{align}
\end{comment}
\begin{align}
    SNR = \frac{\frac{1}{n} \sum_{t=0}^{n-1} Q_0^2 \cdot e^{-\frac{(x-ut)^2}{\sigma_y \sigma_z}} e^{-\frac{y^2}{\sigma_y^2}} e^{-2k_{eff} t}}{\frac{1}{n} \sum_{t=0}^{n-1} Q_N^2 \cdot e^{-\frac{(x-x_0-ut)^2}{\sigma_y \sigma_z}} e^{-\frac{(y-y_0)^2}{\sigma_y^2}} e^{-2k_N t}},
    \label{SNR}
\end{align}
where $Q_N$ is the number of noise VOC released, $k_N$ is the effective rate constant of noise.
%\textcolor{blue}{(\ref{SNR})} depicts the SNR for a single signal and power source. 

\begin{comment}
However, this equation can be modified according to the number of signal and noise sources, which is as follows:
\begin{align}
    SNR_{mn} = \frac{\sum_{m=1} ^k P_0}{\sum_{n=1} ^l P_S},
\end{align}
where $SNR_{mn}$ is the SNR for multiple signal and noise sources, and $m$ and $n$ depict the number of signal and noise sources, respectively.
\end{comment}

\begin{comment}
Due to biotic and abiotic stress, a plant releases VOCs in an instantaneous manner. 
Furthermore, it also releases continuous VOCs as its identity or due to metabolic reasons.
However, in the stress experience, the release VOCs are consist of both instantaneous and continuous emission VOCs.
Hence, it becomes difficult for the receiver plant to distinguish between those two as the received VOCs are in a blended form.
During stress, it is important to the receiver tree to assess the induced stress to the transmitter tree, and in that scenario, the continuously released VOCs will act as noise to the stress-driven VOCs.
The noise model analysis is performed accordingly.
The Signal to Noise Ratio (SNR) can be given by:
\begin{align}
    SNR(f) = \frac{Q_f}{Q_m} \frac{e^{(k_{eff}-i2 \pi f )\frac{x}{u}} \cdot e^{(k_{eff}- i 2 \pi f)^2\frac{\sigma_y \sigma_z}{u^2}}}{e^{-k_{eff}\frac{x}{u}} e^{-\frac{f^2}{2 \Delta f^2}}},
\end{align}
The channel capacity can be expressed as:
\begin{align}
    C = \int_{f1} ^{f2} (1 + SNR(f)) df
\end{align}
\end{comment}

\section{Reception Model}
\label{sec:reception}
There are several mathematical models that exist for the description of the reception process, namely a linear regression model \bluecite{bacci1990bioconcentration}, a partition model \bluecite{riederer1990estimating}, and a mechanistic ordinary differential equation (ODE) based model \bluecite{trapp1995generic}. However, a comparative study of these models shows that the ODE-based model best resembles the experimental data \bluecite{collins2010modeling}. 
\textcolor{black}{Moreover, the ODE formulation explicitly captures the transient air-to-leaf mass exchange process, incorporates parameters such as stomatal conductance and metabolic decay, which enables dynamic system analysis.}
Hence, we consider the ODE-based approach in this study to explore the reception process.

The mass balance approach is followed to determine the deposition of molecules on the leaf. The rate of change in concentration in the leaf is given by \bluecite{trapp1995generic, collins2010modeling}:
\begin{align}
    \frac{dC_R(t)}{dt} = - l C_R(t) + b,
    \label{Recep}
\end{align}
where $C_R(t)$ is the concentration of molecules in the leaf, $l$ is the loss term, and $b$ is the uptake term. The loss term $l$ depends on the surface area of a leaf ($A_l$), conductance ($G_l$), leaf volume ($V_l$), leaf-air partition coefficient ($K_{LA}$), and a pseudo-first-order rate constant for plant growth ($P_{growth}$) in the following way:
\begin{align}
    l = K_{LA}\frac{A_l G_l}{V_l} + P_{growth},
\end{align}
The uptake term ($b$) is given as:
\begin{align}
    b = \frac{A_l G_l}{V_l} c(\vec{r}, t),
\end{align}
where $c(\vec{r}, t)$ is the VOC concentration in the air.
The Fourier transform of (\ref{Recep}) \textcolor{black}{with respect to time `$t$'} can be given as:
\begin{align}
    C_r (f) = \frac{b}{j2\pi f + l} c(\vec{r}, f),
\end{align}
Hence, the transfer function ($H_r (f)$) is given by:
\begin{align}
    H_r (f) = \frac{b}{j2\pi f + l},
\end{align}
The normalized gain ($H_r^{norm} (f)$) is expressed as:
\begin{align}
    H_r^{norm} (f) = \frac{l}{\sqrt{4 \pi^2 f^2 + l^2}},
    \label{reception_gain}
\end{align}
%and the phase is given by:
%\begin{align}
 %   \phi_r(f) = tan^{-1} \frac{2 \pi f}{l},
%\end{align}
and the delay is given by:
\begin{align}
    \tau_r(f) = \frac{2 \pi}{l \left(1 + (2 \pi f/l)^2 \right)},
\end{align}

\begin{comment}
If each VOC's production rate is modulated by a distinct carrier frequency $f_i$, the production rate can be expressed as:
\[
%s^{(i)}(t) = A^{(i)}(t) \cos(2\pi f_i t),
S^{(i)}(t) = A^{(i)} \left( 1 + \sum_{n=0}^{N_i} e^{-\frac{(t - nT_i)^2}{2\sigma_i^2}} \right),
\]
where $A^{(i)}(t)$ is the amplitude envelope of the $i$-th VOC's production.

The emitted signal $q_i(t)$ will then occupy a distinct frequency band centered around $f_i$, allowing the total signal $q_{\text{total}}(t)$ to be separated into its components in the frequency domain using:
\[
Q_{\text{total}}(f) = \sum_{i=1}^N Q_i(f),
\]
where $Q_i(f)$ is the Fourier Transform of $q_i(t)$.

To extract the contribution of each VOC, bandpass filters around each $f_i$ can be applied.
\end{comment}

\section{Results and Discussions}
\label{sec:analysis}
In Sec. \textcolor{blue}{\ref{sec:transmission}}, \textcolor{blue}{\ref{sec:channel}}, and \textcolor{blue}{\ref{sec:reception}}, the transmission, channel, and reception processes of interplant VOC communication are modeled.  In this section, a detailed analysis of the communication architecture is described.
In order to understand the transmitter characteristics from an MC perspective, we consider Quercus ilex (Q. ilex) as the transmitter plant. The monoterpenoid emission is considered as the signal for transmission, and the characteristic parameters are given in Table \textcolor{blue}{\ref{tab:voc_parameters}}.

\begin{table}[t!]
\centering
\caption{Physicochemical parameters for various VOC compounds in \textit{Q. Ilex} \bluecite{niinemets2002model, harley2013roles}.}
\label{tab:voc_parameters}
\begin{tabular}{|>{\centering\arraybackslash}p{1.75cm}|>{\centering\arraybackslash}p{1.2cm}|>{\centering\arraybackslash}p{1.7cm}|>{\centering\arraybackslash}p{0.7cm}|>{\centering\arraybackslash}p{0.8cm}|}
\hline
\textbf{Compound} & $k_a$ (s$^{-1}$) & $k_l$ ($\times$10\textsuperscript{-5} s$^{-1}$) & $k_g$ (s$^{-1}$) & $\eta$ \\
\hline
$\alpha$-Pinene      & 0.002459 & 3.37791 & 0.7  & 0.867 \\
$\beta$-Pinene       & 0.002455 & 5.31881 & 0.9  & 0.846 \\
Myrcene              & 0.001565 & 15.281  & 3    & 0.840 \\
Sabinene             & 0.002953 & 0.4044  & 2.5  & 0.629 \\
cis-$\beta$-Ocimene  & 0.001167 & 4.1857  & 1.5  & 0.782 \\
$p$-Cymene           & 0.000844 & 0.7434  & 2    & 0.697 \\
$\gamma$-Terpinene   & 0.001395 & 2.4189  & 1.7  & 0.811 \\
$\alpha$-Terpinolene  & 0.001126 & 1.1959  & 1.3  & 0.736 \\
$\beta$-Phellandrene  & 0.001319 & 0.5852  & 2.3  & 0.762 \\
\hline
\end{tabular}
\end{table}

\begin{figure}[t!]
    \centering
    \includegraphics[width=\linewidth]{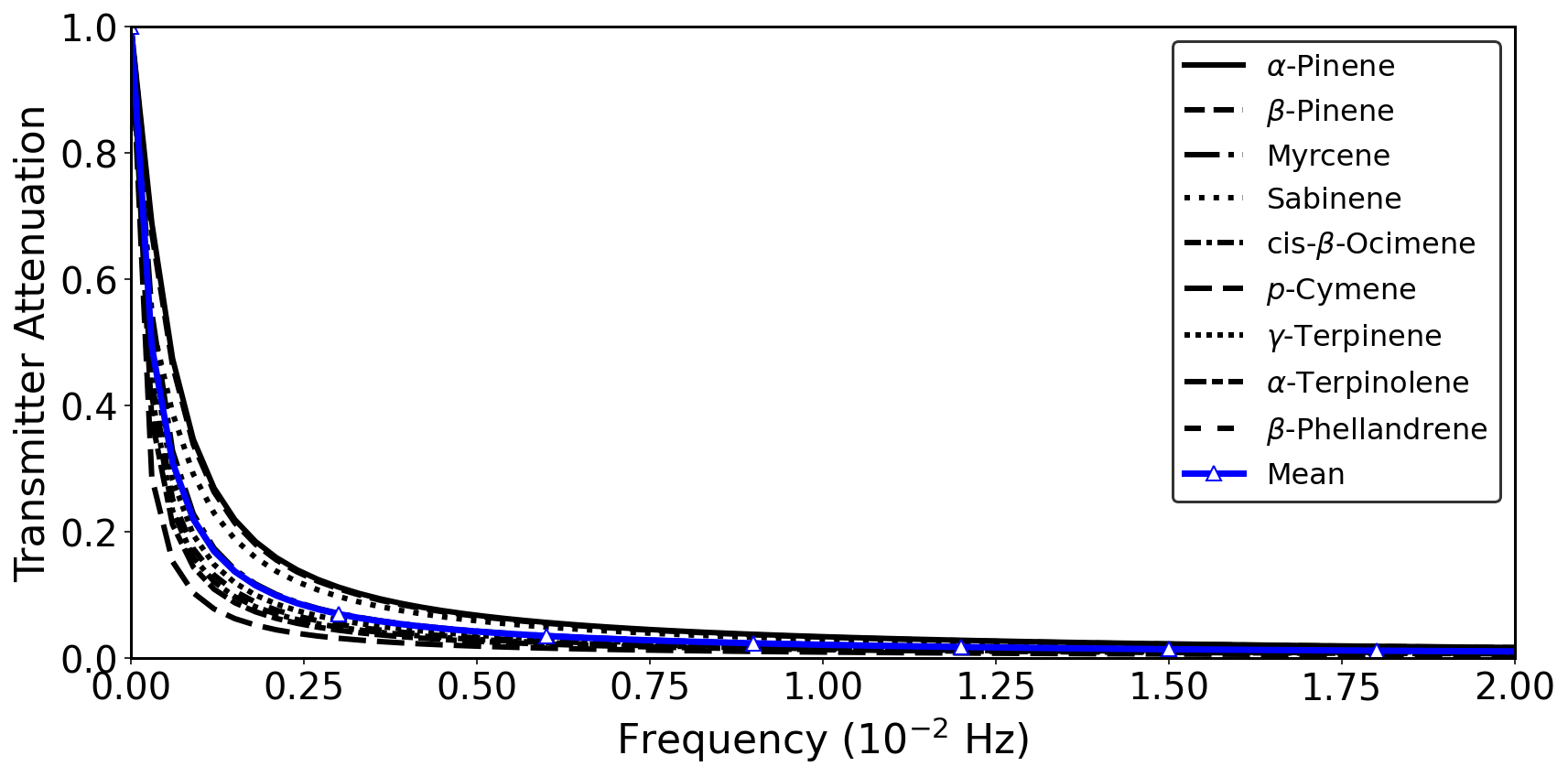}
    \caption{Attenuation of transmission process over frequency.}
    \label{fig:tx_atte}
\end{figure}
\begin{figure}[t!]
    \centering
    \includegraphics[width=\linewidth]{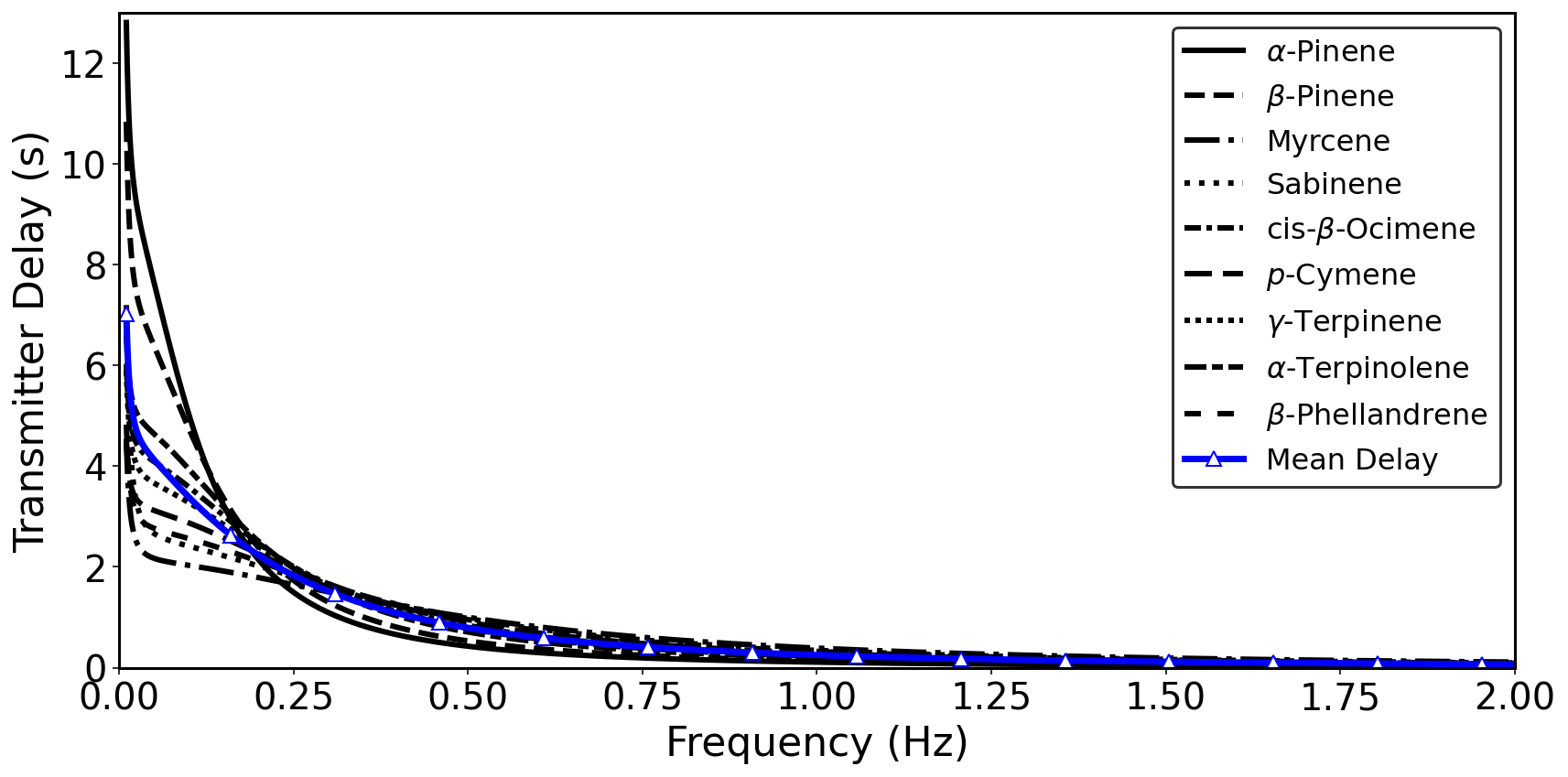}
    \caption{Delay of transmission process with respect to frequency.}
    \label{fig:tx_del}
\end{figure}

\textcolor{black}{In this study, frequency represents the temporal dynamics of VOC signals. 
Low-frequency components correspond to slowly varying signals, like the circadian rhythm, whereas high-frequency components represent rapid 
fluctuations in VOC release.
Thus, the frequency response characterizes which temporal window of VOC emissions can 
be effectively received by the receiver in interplant communication.}
\textcolor{black}{
%The frequency responses indicate VOC-specific temporal attenuation. 
In the proposed model, each VOC propagates independently without cross-VOC coupling, and at the receiver, VOC-specific uptake mechanisms enable effective demultiplexing based on chemical identity.}

Firstly, the attenuation of the transmission process is illustrated in Fig. \textcolor{blue}{\ref{fig:tx_atte}}. 
The release of blended monoterpenoids from the leaves constitutes several VOCs, and it can be seen that the attenuation drops sharply over frequency. 
%Even though stress-driven VOC emission is modeled as an instantaneous burst, which corresponds to a broad frequency spectrum, 
Due to biological constraints like VOC transport through the cuticle and chemical partitioning, higher frequencies are highly attenuated.
This results in an effective bandwidth of less than 0.002 Hz, which resembles a narrowband, Gaussian-approximated delta function.
Fig. \textcolor{blue}{\ref{fig:tx_del}} depicts the delay of the transmission process over the frequency. 
Similar to the attenuation, the delay profile of different VOCs also decreases with frequency. 
In Q. ilex, $\alpha$-pinene is the most abundant monoterpenoid, and its delay is maximum initially, which is 12.8 sec, though the delay doesn't depend upon the number of released molecules, whereas the mean delay of the released VOCs is 7.02 sec.
The observed sharp decrease in delay suggests that leaf cells allow fast VOC movements, which enable signal propagation for high frequencies with zero delay.

\begin{comment}
\begin{table}[t!]
    \centering
    \caption{Parameters}
    \begin{tabular}{c|c}
        Distance & 10, 20, 50, 100, 200, 500 \\
        Wind speed  & 3, 4, 5, 6, 7\\
        $Q_0$ & 10000 \\
        OH & $2\times 10^6$ molecules/cm\textsuperscript{3} \bluecite{atkinson2000atmospheric}\\
        O\textsubscript{3}& $7 \times 10^{11}$ molecules/cm\textsuperscript{3} \bluecite{logan1985tropospheric}\\
        NO\textsubscript{3}& $1 \times 10^{10}$ molecules/cm\textsuperscript{3} \bluecite{atkinson2000atmospheric}\\
    \end{tabular}
    \label{tab:my_label}
\end{table}
\end{comment}
\begin{table}[t!]
\centering
\caption{VOC compositions in Q.Ilex and Pinus pinea.}
\label{tab:ratio}
\begin{tabular}{|>{\centering\arraybackslash}p{2cm}|>{\centering\arraybackslash}p{2cm}|>{\centering\arraybackslash}p{2.6cm}|}
\hline
\textbf{Compound} & \textbf{Composition in Q. ilex (\%)} \bluecite{niinemets2002model} & \textbf{Composition in Pinus pinea (\%)} \cite{noe2006emissions} \\
\hline
$\alpha$-Pinene & 32.0718 & 0.4341 \\
$\beta$-Pinene & 24.6972 & 0.0894 \\
Myrcene & 12.9217 & 1.3788 \\
Sabinene & 10.5861 & 0.4979 \\
cis-$\beta$-Ocimene & 5.1687 & 1.3660 \\
$p$-Cymene & 4.3577 & --- \\
$\gamma$-Terpinene & 4.2604 & --- \\
$\alpha$-Terpinolene & 3.2115 & --- \\
$\beta$-Phellandrene & 2.7249 & --- \\
Acetone & --- & 12.3057 \\
Limonene & --- & 1.2384 \\
trans-$\beta$-Ocimene & --- & 70.7136 \\
Linalool & --- & 6.4343 \\
\hline
\end{tabular}
\end{table}

 \begin{figure}[t!]
    \centering
    \includegraphics[width=\linewidth]{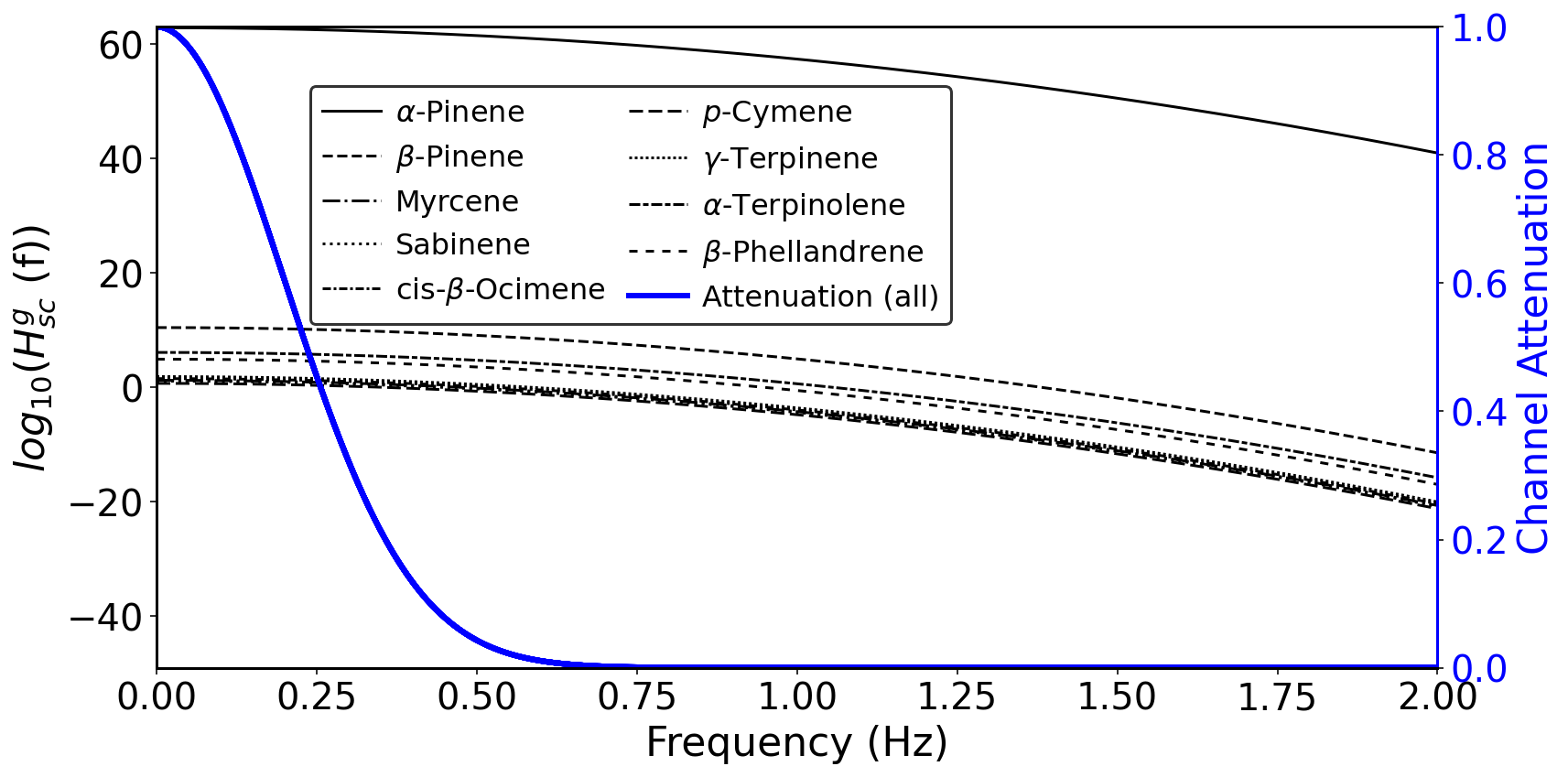}
    \caption{$log_{10} (H_{sc}^g (f))$ and attenuation of the channel over frequency at a distance of 100 meters with a velocity of 7 $m/s$.}
    \label{fig:log10 and VOCs attenuation}
\end{figure}   
The interplant channel characteristics for stress-driven or instantaneous VOC release are modeled in Sec. \textcolor{blue}{\ref{subsec:puff}} and are analyzed here.
For the propagation of blended VOCs in the channel, a total of $Q_0$ molecules is considered. 
The number of individual VOC released by Q. ilex is followed according to the ratio of the total emission pool, which is given in Table \textcolor{blue}{\ref{tab:ratio}}.
The rate constants that are required to determine $k_{eff}$ are given in Table \textcolor{blue}{\ref{tab:rate_cons}}.
Moreover, the concentration of OH, O\textsubscript{3}, and NO\textsubscript{3} are 2$\times$10\textsuperscript{6}, 7$\times$10\textsuperscript{11}, and 1$\times$10\textsuperscript{10} molecules/cm\textsuperscript{3}, respectively, and the number of released VOCs ($Q_0$) is 10000.

Fig. \textcolor{blue}{\ref{fig:log10 and VOCs attenuation}} depicts the channel behavior both in terms of gain and attenuation at a distance of 100 meters with a velocity of 7 $m/s$.
It can be observed that the gain, plotted as $log_{10} (H_{sc}^g (f))$, widely varies according to the type of VOC and decreases gradually with the frequency.
This variation arises due to the differences in the kinetic parameters of different VOCs.
However, the attenuation for the channel is identical across all VOCs, which rapidly drops as the frequency increases.
It confirms that individual VOCs may behave differently due to molecular properties, but the medium imposes a common high-frequency attenuation, which significantly limits the effective bandwidth.
%\textcolor{blue}{The frequency responses in Fig. \textcolor{blue}{\ref{fig:log10 and VOCs attenuation}} indicate VOC-specific temporal attenuation.  In the proposed model, each VOC arrives at the receiver as a distinct concentration input to its corresponding uptake dynamics, which ensures the separability of the transported VOCs.}

\begin{table}[t!]
\centering
\caption{Arrhenius rate constant ($k$ (cm\textsuperscript{3} molecule\textsuperscript{-1} s\textsuperscript{-1} )) at room temperature.}
\begin{tabular}{|c|c|c|c|c|}
\hline
\textbf{VOC compound} & \makecell{OH \\ ($\times 10^{-12}$)} & \makecell{NO\textsubscript{3} \\ ($\times 10^{-11}$)} & \makecell{O\textsubscript{3} \\ ($\times 10^{-17}$)} & Ref. \\ 
\hline
$\alpha$-pinene & 52.3 & 84 & 6.16 & \bluecite{atkinson2003atmospheric}\\
$\beta$ pinene  & 74.3  & 15 & 2.51 & \bluecite{atkinson2003atmospheric}\\
 Myrcene & 215 & 1.1 & 47 & \bluecite{atkinson2003atmospheric}\\
 Sabinene & 117 & 1.0 & 8.3 & \bluecite{atkinson2003atmospheric}\\
 cis-$\beta$-Ocimene & 252 & 2.2 & 54 & \bluecite{atkinson2003atmospheric} \\
 p-Cymene & 151 & 0.99 & 0.2 & \bluecite{corchnoy1990kinetics, atkinson1990rate} \\
 $\gamma$-Terpinene & 177 & 2.9 & 14 & \bluecite{atkinson2003atmospheric}\\
 $\alpha$-Terpinolene & 225 & 9.7 & 190 & \bluecite{atkinson2003atmospheric}\\
 $\beta$-Phellandrene & 168 & 8 & 4.7 & \bluecite{atkinson2003atmospheric}\\
 Acetone & 0.17 & 2.9$\times$10\textsuperscript{-6} & 0.9$\times$10\textsuperscript{-4} & \bluecite{atkinson2003atmospheric} \\
 Limonene & 164 & 1.22 & 21 & \bluecite{atkinson2003atmospheric}\\
 trans-$\beta$-Ocimene & 252 & 2.2 & 54 & \bluecite{atkinson2003atmospheric}\\
 Linalool & 159 & 1.12 & 43 & \bluecite{atkinson1995rate}\\
 \hline
\end{tabular}
\label{tab:rate_cons}
\end{table}

\begin{figure}[t!]
    \centering
    \includegraphics[width=\linewidth]{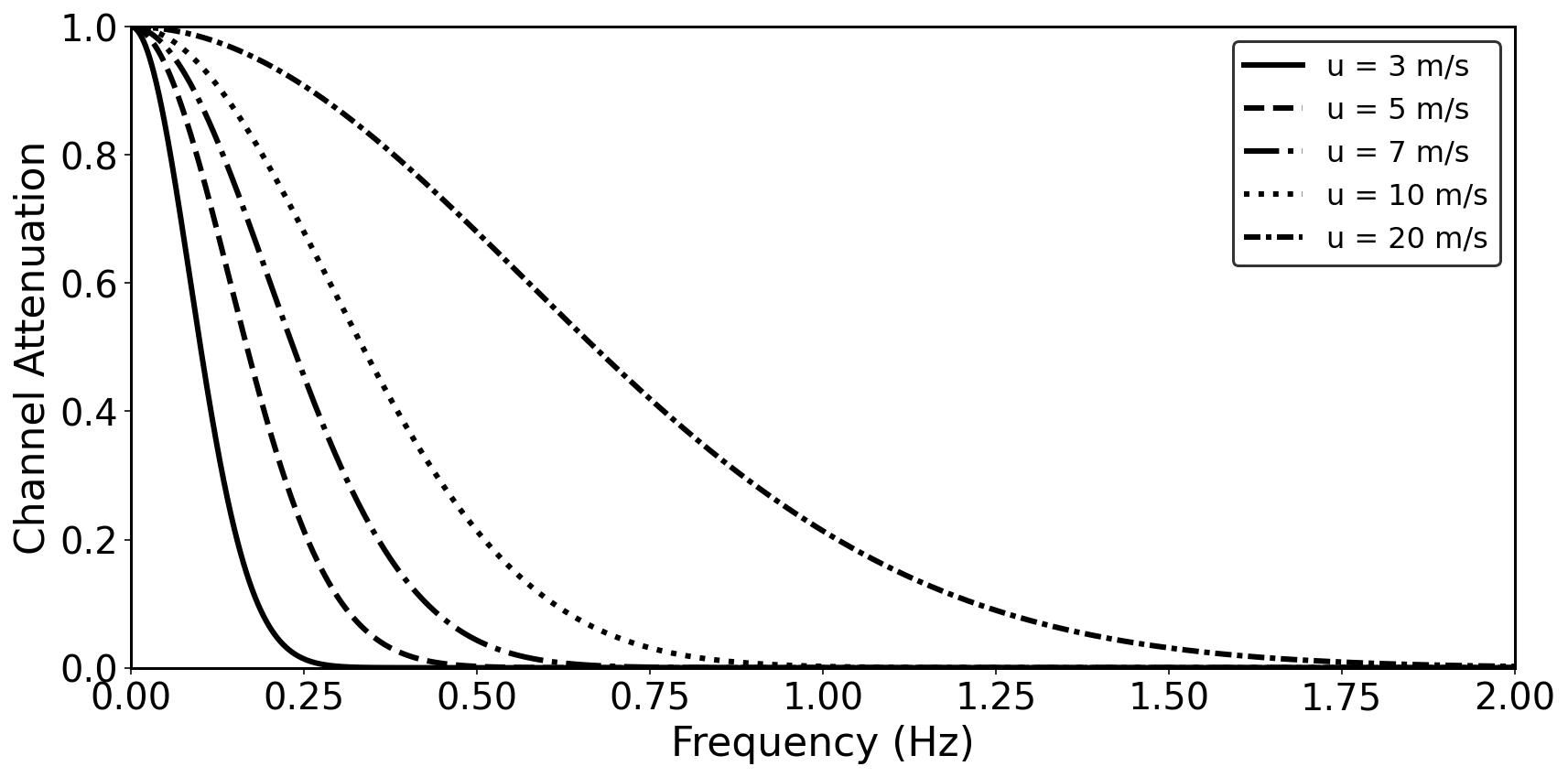}
    \caption{Channel attenuation with respect to frequency for different velocities at a distance of 100 meter.}
    \label{fig:velocity}
\end{figure}
Fig. \textcolor{blue}{\ref{fig:velocity}} depicts the effect of wind velocity variation on the channel attenuation over frequency at a fixed distance of 100 meters. 
%The channel exhibits low-pass behavior for all velocities, with higher attenuation at increasing frequencies.
It is noticeable that as the wind velocity increases, the attenuation shifts toward higher frequencies. 
This means that higher velocities result in broader bandwidths.
%, and high-frequency components of the VOC signal can propagate more effectively
This trend can be explained by the fact that increasing the wind speed reduces the presence time of VOCs in the medium, which decreases the effect of reactive decay and dispersion. 
%Hence, the attenuation pattern confirms that the wind speed plays a crucial role in shaping the frequency response of the VOC-based MC channel.
\begin{figure}[t!]
    \centering
    \includegraphics[width=
    \linewidth]{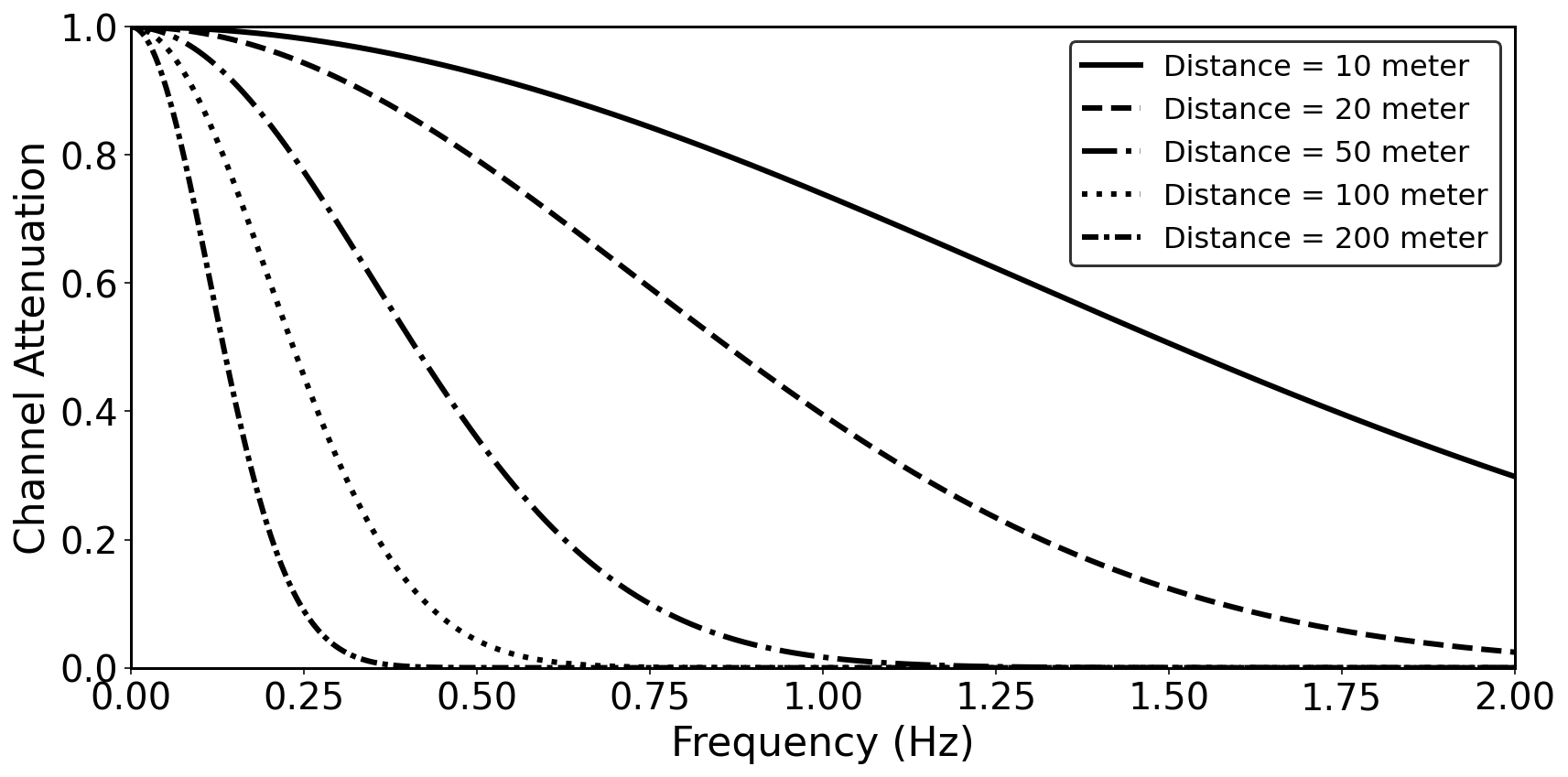}
    \caption{Channel attenuation with respect to frequency at different distances at a velocity of 7 $m/s$.}
    \label{fig:distance}
\end{figure}
Fig. \textcolor{blue}{\ref{fig:distance}} depicts the variation of channel attenuation over frequency for different transmission distances with a fixed wind velocity of 7 $m/s$. 
%It illustrates that as the distance increases, the attenuation accelerates, which leads to a reduced effective bandwidth for greater distances. 
At shorter distances (e.g., 10 or 20 meters), the attenuation is gradual, allowing a broader spectrum of frequency components to pass through.
The physical interpretation of this behavior is due to VOCs undergoing more interactions with the atmospheric reactants and experiencing a greater spread, which weakens the signal strength at higher frequencies.
%This figure explains that transmission distance is a limiting factor for interplant VOC communication and must be considered while designing a bio-inspired MC system.

\begin{comment}
\begin{figure}[t!]
    \centering
    \includegraphics[width=\linewidth]{figures/VOC ateneuation&log.png}
    \caption{Caption}
    \label{fig:enter-label}
\end{figure}

\begin{figure}[t!]
    \centering
    \includegraphics[width=\linewidth]{figures/multiple_transmitter_channel.png}
    \caption{at u = 20 $m/s$}
    \label{fig:enter-label}
\end{figure}
\end{comment}

\begin{figure}[t!]
    \centering
    \includegraphics[width=\linewidth]{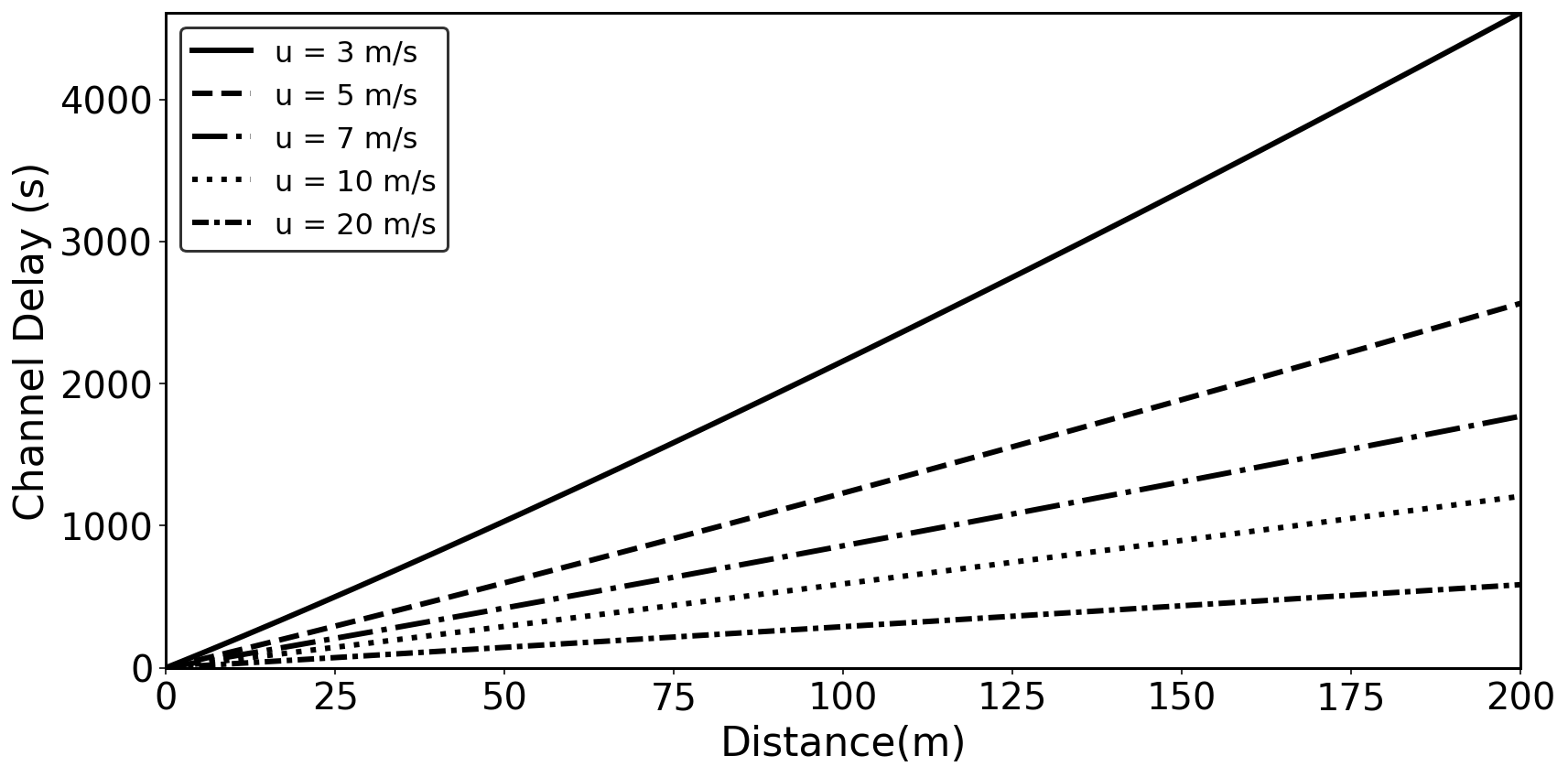}
    \caption{Delay of channel with respect to distance at different velocities.}
    \label{fig:cx_delay}
\end{figure}
Fig. \textcolor{blue}{\ref{fig:cx_delay}} illustrates the variation of channel delay with respect to distance for different wind velocities.
The graph shows a linear relationship between delay and distance, which indicates that the channel introduces uniform propagation delay across distances.
It is also observed that the higher velocities lead to significantly lower delays at the same distance. 
%For instance, at 200 meters, the delay is 4609 seconds for 3 $m/s$, but drops to 584 seconds for 20 $m/s$.
This behavior can be explained by the fact that faster wind helps VOCs to propagate more quickly, reducing the time required to reach the receiver plant. 
%The separation between the curves also increases with distance, which emphasizes the impact of wind speed over longer ranges.
Hence, for faster and more reliable VOC-based information transfer, higher velocities, short distances, or both are favorable.

\begin{figure}[t!]
    \centering
    \includegraphics[width=\linewidth]{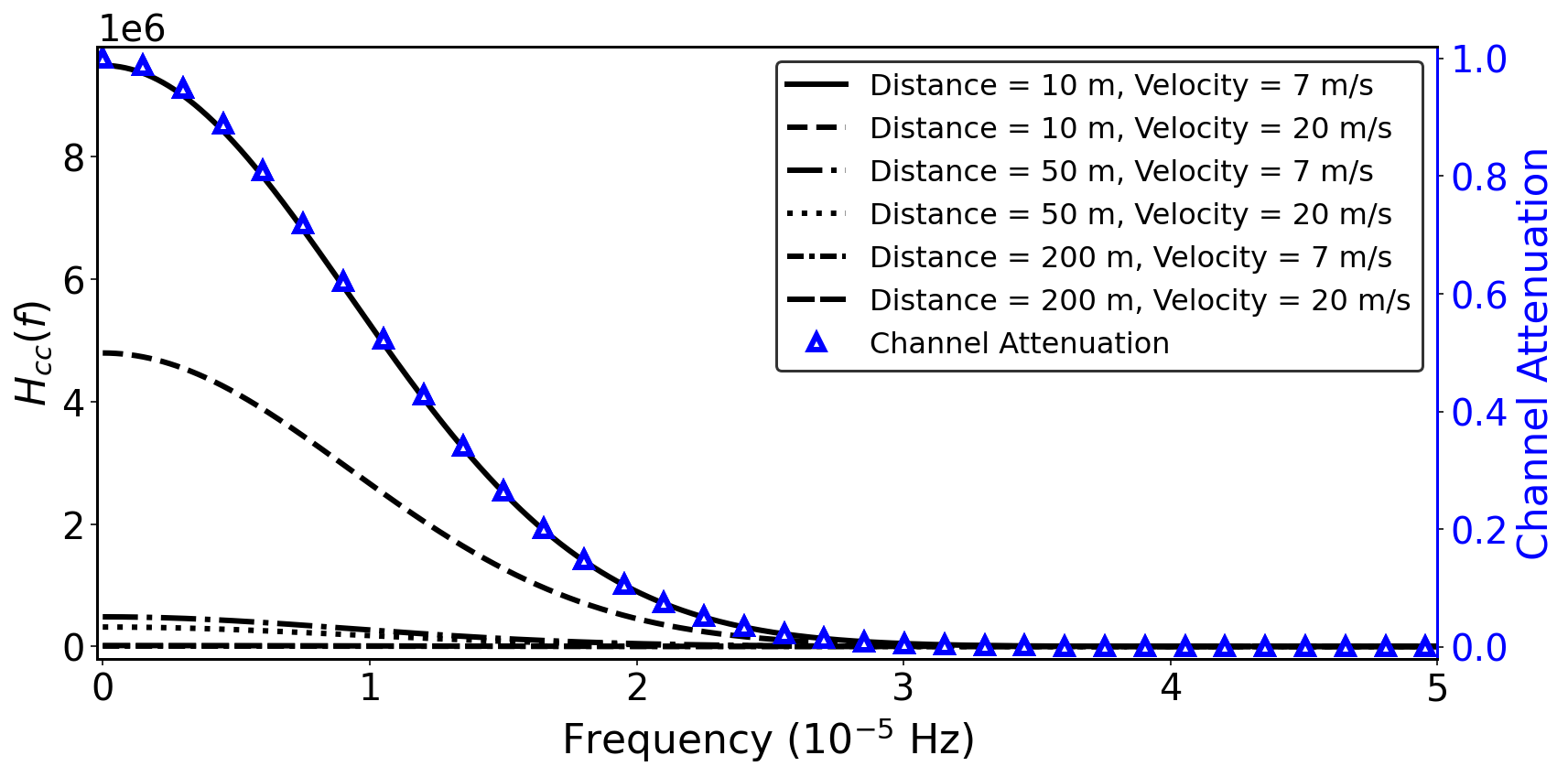}
    \caption{$H_{cc}^n(f)$ and channel attenuation with respect to frequency at different distances and velocities for constitutive VOC emission.}
    \label{fig:plumel}
\end{figure}

\begin{comment}
\begin{figure}[t!]
    \centering
    \includegraphics[width=\linewidth]{figures/bandwidth_1.png}
    \caption{Bandwidth of the channel with respect to distance at different velocities. }
    \label{fig:bw}
\end{figure}
\end{comment}
%Constitutive VOC release, which is modeled in Sec. \textcolor{blue}{\ref{subsec:plume}} is described hereafter.
Fig. \textcolor{blue}{\ref{fig:plumel}} illustrates the
gain $H_{cc}(f)$ and channel attenuation with respect to frequency for various combinations of propagation distance and wind velocity for continuous VOC release conditions.
%It is observed that the channel exhibits a strong low-pass behavior, where the gain sharply declines with increasing frequency. 
The plot shows that shorter distances and lower velocities yield higher gain with respect to frequency. 
This is because slower wind speeds allow VOCs to remain concentrated in the channel for a longer period, which minimizes molecular spread and degradation. 
The normalized attenuation (shown by the blue triangular curve) falls off rapidly, implying that the channel can only support very low-frequency components, which are typically under 2 $\times 10^{-5}$ Hz. 
%This suggests that only very slowly varying biochemical signals can reliably propagate, which is indeed the physicality of the constitutive VOC release mechanism of plants for MC.
%Fig. \textcolor{blue}{\ref{fig:bw}} elucidates the variation of channel bandwidth with respect to distance for different wind velocities. It is observed that the effective bandwidth decreases sharply with increasing distance. For shorter distances and higher velocities, the bandwidth remains significantly high due to low dispersion and VOC degradation. 

\begin{figure}[t!]
    \centering
    \includegraphics[width=0.9\linewidth]{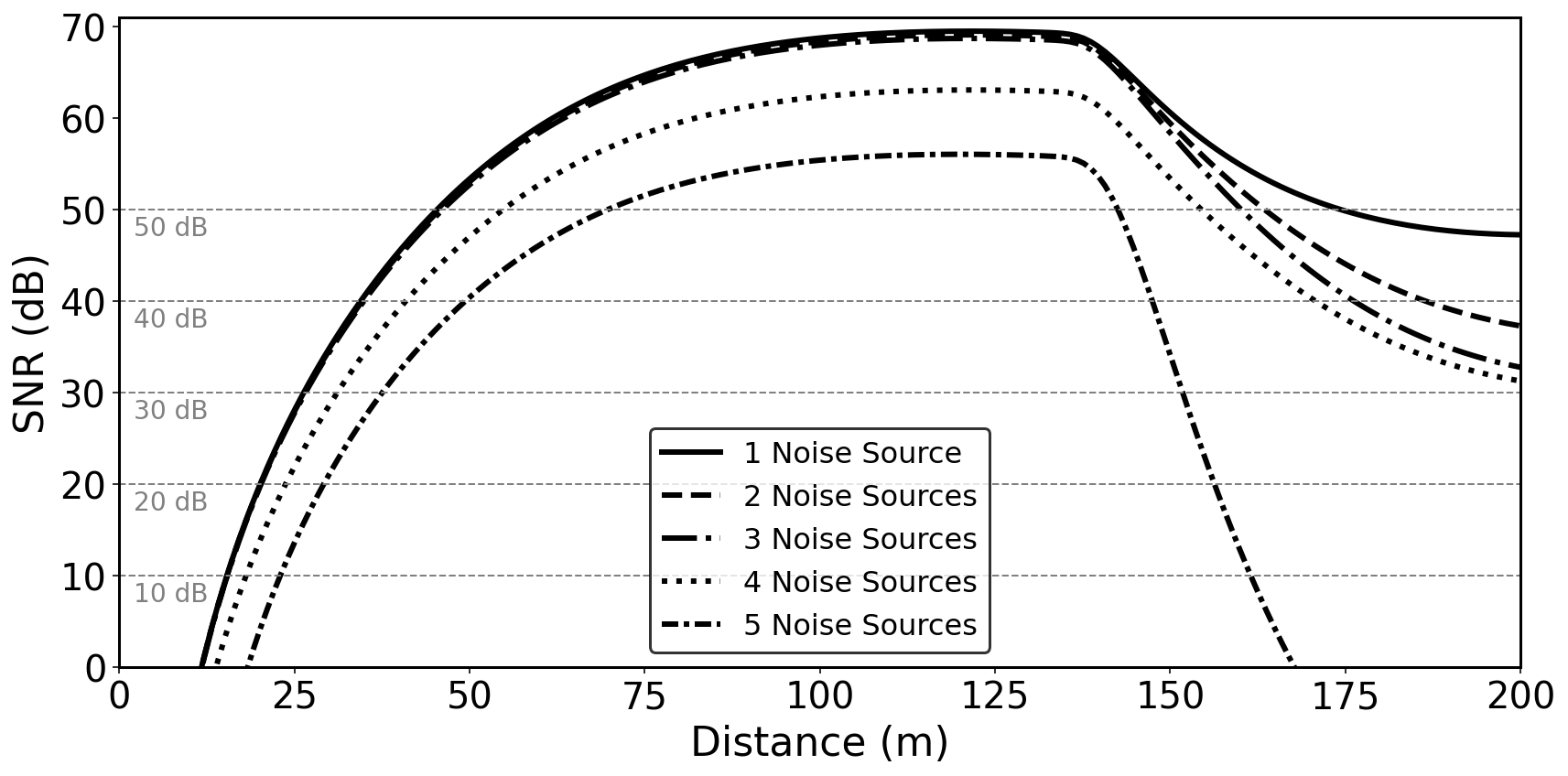}
    \caption{SNR with respect to distance considering 5 nearby noise sources at 7 $m/s$.}
    \label{fig:SNR}
\end{figure}

%Furthermore, the noise analysis that is modeled in \textcolor{blue}{\ref{noise}} is analyzed from the ICT perspective.
Fig. \textcolor{blue}{\ref{fig:SNR}} depicts the SNR as a function of distance at a velocity of 7 $m/s$ under the influence of the increasing number of nearby noise sources up to five. 
In total, five noises are considered, and their positions are (2,1), (2,-1), (4,2), (4,-2), and (6,0).
The noise sources are analyzed as a cumulative source.
\textit{Pinus pinea} is considered as a noise source; due to different leaf architecture and species, the emission of this widely differs from that of \textit{Q. ilex}, as mentioned in Table \textcolor{blue}{\ref{tab:ratio}}.
\textcolor{black}{As shown in Table \textcolor{blue}{\ref{tab:ratio}}, VOC emissions from different species exhibit both shared and species-specific VOCs. 
Several VOCs like $\alpha$-pinene, $\beta$-pinene, myrcene, etc., are present in both Q.ilex and Pinus pinea along with significantly different VOCs.
This partial overlap in VOCs indicates that emissions from different species may not be fully distinguishable at the receiver, which leads to potential interference.}
%It is notable that the SNR is highly influenced by the position of the noise and the number of noise sources.
With this step-up, the SNR initially increases, reflecting the accumulation of signal as it travels through the medium.
However, beyond approximately 140 meters, SNR begins to decline or flatten depending on the noise configuration.
This behavior arises as the VOC signal is modeled as a Gaussian pulse centered at $x=u \cdot t$, which has a finite window, considering $t \in [0, 20]$ seconds and $u= 7$ $m/s$.
%For this reason, if the signal front reaches $x \approx 140$ m, the signal energy starts to decay, but the noise continues to contribute to the background concentration.
The variation between SNR curves is due to different noise sources either constructively or destructively interfering with the signal. 
%This result highlights the importance of considering the spatiotemporal structure of signal propagation for long-distance VOC communication.
%With the noise analysis, it is observed that the VOC communication architecture can be used for different purposes depending on the interplant distance.
%The SNR below 10 dB indicates that the signal is hard to distinguish from noise and is not suitable for any kind of communication. 
%SNR between 10 and 20 dB is of poor quality, but SNR from 20 to 30 dB is considered in low-speed communication.
%30 to 40 dB SNR is used in voice communication and cell phones.
%Finally, 
An SNR above 40 dB is considered to be of excellent quality and is used in digital signals and high-speed communication systems.
Consequently, based on the obtained SNR, VOC-based MC communication can link the nano-network with micro-scale devices, which can be merged with traditional communication through the internet in the micro-device.

\begin{figure}[t!]
    \centering
    \includegraphics[width=0.9\linewidth]{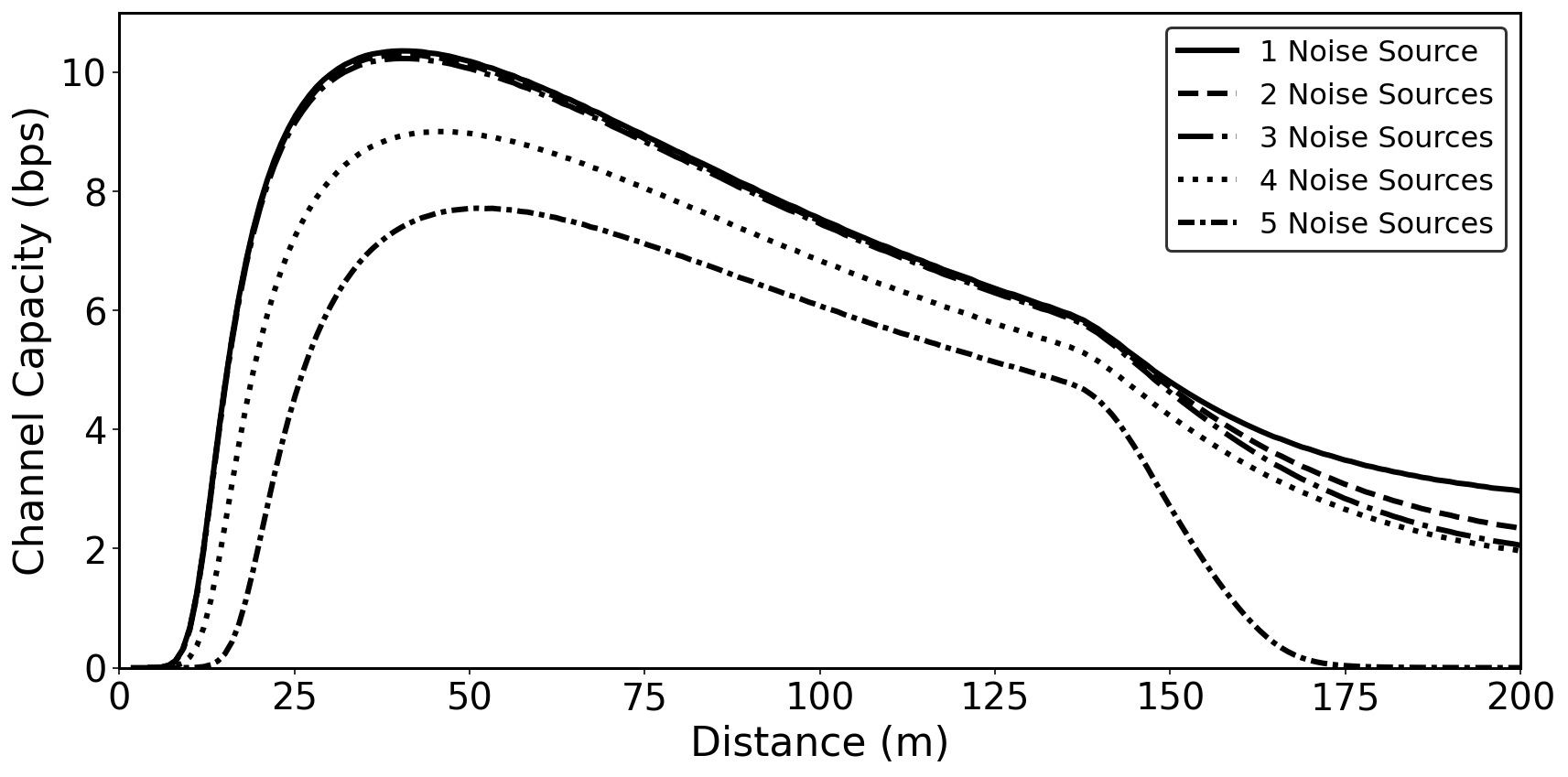}
    \caption{Channel capacity with respect to distance at 7 $m/s$ considering 5 nearby noise sources.}
    \label{fig:capacity}
\end{figure}
Fig. \textcolor{blue}{\ref{fig:capacity}} illustrates the channel capacity calculated using the Shannon formula, incorporating the SNR profile from Fig. \textcolor{blue}{\ref{fig:SNR}} and bandwidth. 
The Shannon capacity ($C$) formula is expressed as:
\begin{align}
    C = B \cdot log_2 (1+SNR),
\end{align}
where $B$ is the bandwidth, and $C$ is expressed in bits per second (bps).
\textcolor{black}{Shannon capacity provides a fundamental upper bound on the maximum achievable information rate over a noisy communication channel, which has been widely used in MC to characterize diffusion-based channels under stochastic propagation and interference \bluecite{bicen2018shannon}.
Since the VOC-based interplant communication follows advection-diffusion dynamics, Shannon capacity provides an information-theoretic framework to evaluate the communication system.}
It is observed that the capacity initially increases with distance, reaching a peak before gradually declining. 
%This can be explained by a trade-off between the initially increasing bandwidth at short ranges and the declining SNR at longer distances. 
%At higher distances, both bandwidth and SNR decrease, resulting in a capacity drop. 
Furthermore, increasing noise levels considerably lowers the overall channel capacity.
These results collectively indicate that the MC physical channel layer in the interplant communication system is highly sensitive to propagation distance, flow conditions, and background noise. 
Optimization of these parameters is required to ensure high-throughput and reliable information transfer in interplant communication networks.

\begin{figure}[t!]
    \centering
    \includegraphics[width=0.9\linewidth]{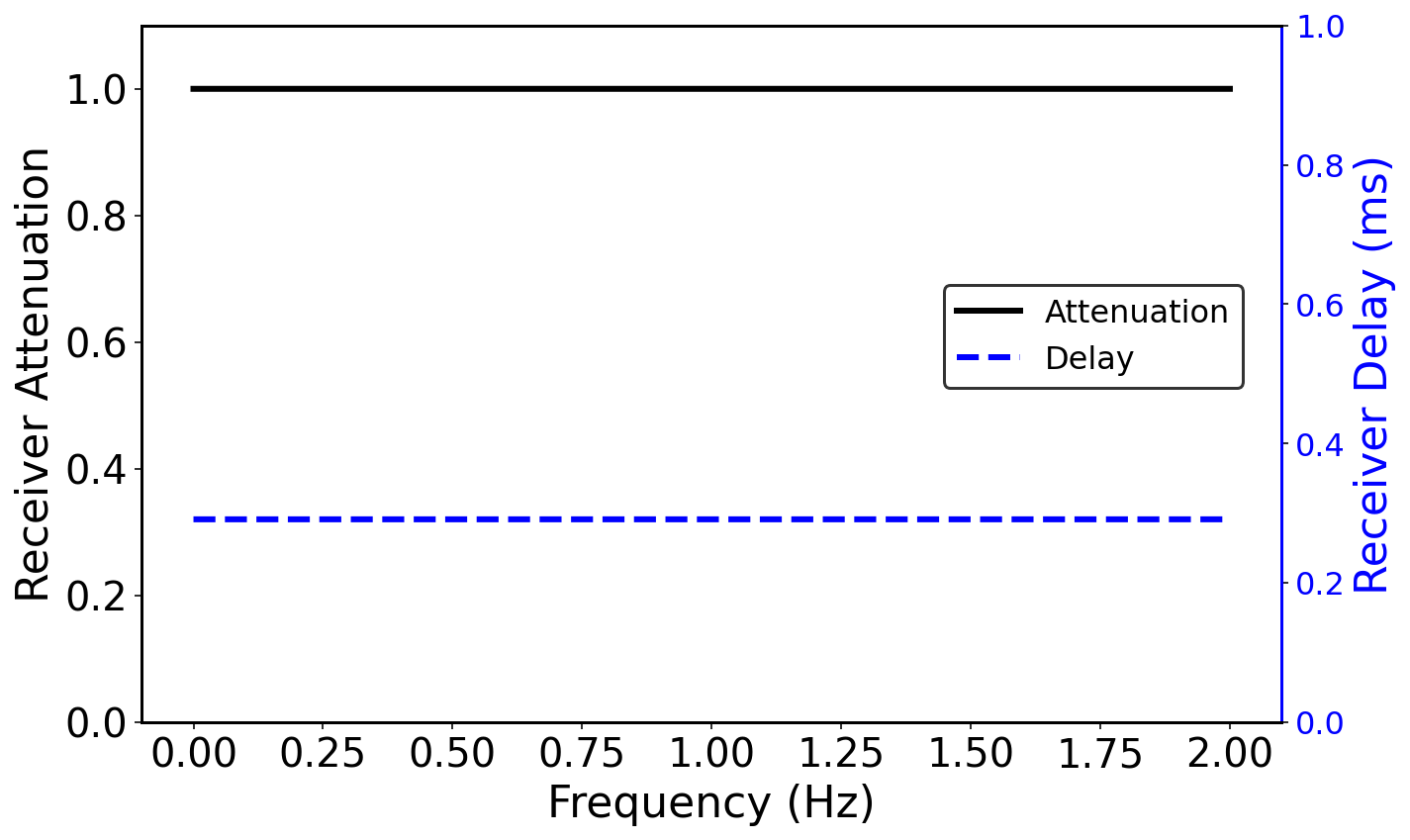}
    \caption{Attenuation and delay of receiver over frequency.}
    \label{fig:rx_g_d}
\end{figure}
Fig. \textcolor{blue}{\ref{fig:rx_g_d}} shows the attenuation and delay of the receiver over frequency by considering $A_l = 5$ m\textsuperscript{2}, $G_l=86.4$ m/day, $V_l=0.002$ m\textsuperscript{3}, $K_{LA}=10$, $P_{growth}=0.035$ /day \bluecite{collins2010modeling,terzaghi2021microbial}.
The attenuation of the receiver remains constant at $\sim$1, and the delay stays flat at $\sim$0.3 ms. 
This indicates that the plant receiver exhibits a frequency-independent response, which resembles a linear and memoryless communication channel.

Furthermore, an end-to-end channel model can be done by modeling end-to-end channel attenuation ($H_e(f)$) with the obtained normalized gain in \textcolor{blue}{(\ref{NGT})}, \textcolor{blue}{(\ref{norm_gain})}, and \textcolor{blue}{(\ref{reception_gain})} via $H_e(f) = H_{Tx}^n (f) \cdot H_{sc}^n (f) \cdot H_r^n(f)$. 
It can be inferred from Fig. \textcolor{blue}{\ref{fig:tx_atte}}, \textcolor{blue}{\ref{fig:velocity}}, and \textcolor{blue}{\ref{fig:rx_g_d}} that the end-to-end channel attenuation is dominated by the transmitter attenuation as the attenuation rapidly drops to zero at $\sim$10 mHz, which implies that the end-to-end channel has a very narrow bandwidth.
Hence, it can only support very slow varying or low-frequency signaling.
%Although the end-to-end channel is limited by the transmitter's characteristics, the physical channel exhibits a broader bandwidth, implying that the physical channel is capable of transmitting faster variations in VOC concentration.
\begin{figure}[t!]
    \centering
    \includegraphics[width=0.9\linewidth]{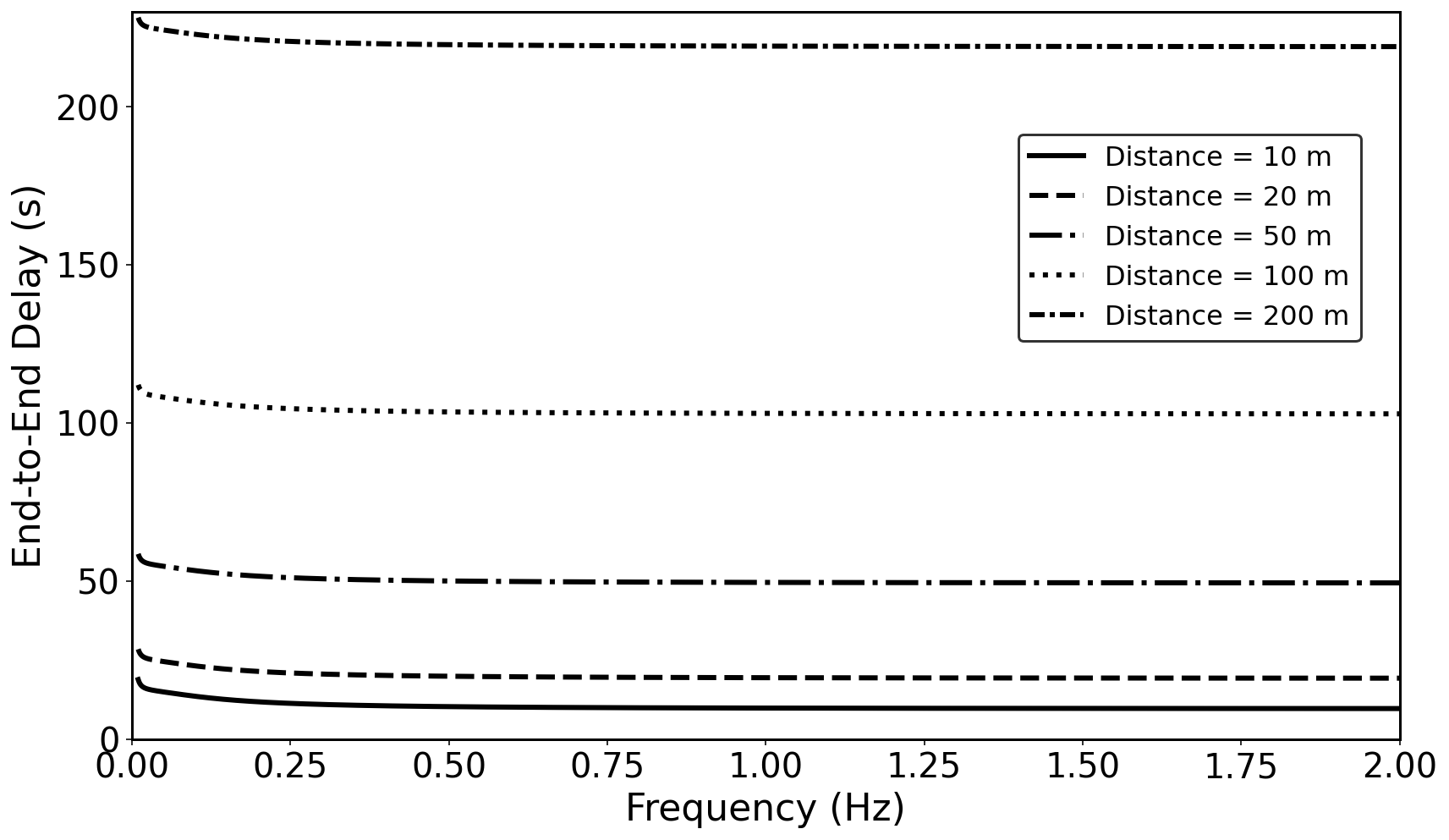}
    \caption{End-to-end channel delay over frequency at different transmitter-receiver distance.}
    \label{fig:e-t-e-delay}
\end{figure}
Moreover, the delay of the end-to-end channel can be given by $\tau_e(f) = \tau_{Tx} (f) + \tau_{sc} (f) + \tau_{r} (f)$.
Fig. \textcolor{blue}{\ref{fig:e-t-e-delay}} depicts the delay of the end-to-end channel at a constant wind velocity of 7 $m/s$ for multiple transmitter-receiver distances. 
%It can be observed that the delay is dominated by the physical channel delay.
The delay increases with increasing distance due to the longer time required for VOCs to drift through the channel, which is dominant over VOC diffusion in the physical channel.

\section{Conclusion}
\label{sec:conclusion}
In this paper, an end-to-end VOC-based interplant MC channel is studied from ICT perspective.
%In this paper, the VOC-based interplant MC channel is studied from the ICT perspective.
%It is comprised of three stages: transmission, physical channel, and reception.
%In the transmission process, leaves produce VOCs and store them in aqueous and lipid pools; from there, those VOCs diffuse to the intracellular air space and are released into the ambient air.
%In the physical channel, these VOCs travel considering the advection-diffusion mechanism in the turbulent wind flow condition, where they undergo VOC degradation due to the presence of oxidants.
%The propagation of the VOCs can be modeled in different ways according to the emission type; for the stress-induced VOC emission, the Gaussian puff model is preferred, whereas the Gaussian plume model is used for constitutive VOC emission.
%Lastly, when these VOCs reach the receiver plant, the VOCs are uptaken by the leaves, and physiological signaling pathways are created according to the received VOC concentration or the ratio in the blended VOC.
\textcolor{black}{This study examines the impact of distance and flow velocity on 
VOC-based interplant MC in the presence of VOC degradation and demonstrates how these parameters influence attenuation, delay, and frequency-dependent behavior, thereby determining the effective communication 
range and reliability.}
\textcolor{black}{The results exhibit that the end-to-end VOC propagation channel shows low-pass characteristics due to the biological emission patterns, which are typically governed by slow processes such as stress responses and circadian rhythms.}
\textcolor{black}{Despite the biologically imposed bandwidth limitation, atmospheric advection-diffusion enables passive VOC transport over a large spatial range during propagation, which makes it fundamentally suited for sparse, long-range broadcast signaling of stress events.}
\textcolor{black}{The bandwidth limitations may be overcome through multi-dimensional encoding, including VOC type, concentration ratios, and temporal patterns, etc., or cooperative transmission from multiple sources.}
\textcolor{black}{VOC-based interplant MC can enable distributed environmental monitoring and early stress detection in agricultural systems, where plants act as natural transmitters and receivers of VOC signals. Furthermore, the inherent long-range and broadcast nature of VOC signaling can be leveraged for designing energy-efficient, bio-inspired communication networks.
}

\appendices
\section{Derivation of (\textcolor{blue}{6}) from (\textcolor{blue}{1-5})}
\label{A_A}
Considering Fourier transformation of \eqref{s_a}, \eqref{s_l}, and \eqref{s_g} with respect to time `$t$', we obtain:
\begin{align}
    S_a(f) = \frac{\eta}{j2\pi f +k_a} P(f) \tag{1a}\\
    S_l(f) = \frac{1-\eta}{j2\pi f +k_l} P(f) \tag{2a}\\
    S_g(f) = \frac{k_aS_a(f)+k_lS_l(f)}{j2\pi f+k_g} \tag{3a}
\end{align}
Substituting (\textcolor{blue}{1a}), (\textcolor{blue}{2a}), and (\textcolor{blue}{3a}) in \eqref{Qf}, we obtain:
\begin{align}
    \frac{E(f)}{P(f)} = \left ( \frac{\eta k_a}{j2\pi f +k_a} + \frac{(1-\eta) k_l}{j2 \pi f + k_l} \right ) \frac{k_g}{j2\pi f + k_g} \nonumber
    \label{fr_5}
\end{align}
which is expressed as:
\begin{align}
    H_{Tx} (f) & = \frac{k_g}{j2 \pi f + k_g} \times \left(\frac{k_a \eta}{j2 \pi f + k_a} + \frac{k_l (1 - \eta)}{j2 \pi f + k_l} \right), \tag{6}
\end{align}

\section{Derivation of (17) from (16)}
\label{A_B}
\begin{align}
    \frac{\partial c(\vec{r}, t)}{\partial t} + \nabla \cdot (c (\vec{r}, t) \vec{u}) - \nabla \cdot (K \nabla c(\vec{r},t)) = S (\vec{r},t), 
    \label{final2}
    \tag{18}
\end{align}
Considering the wind velocity in the x-direction ($u$,$0$,$0$), \eqref{final2} reduces to:
\begin{align}
    \frac{\partial c}{\partial t} + u \frac{\partial c}{\partial x} = K_x \frac{\partial^2 c}{\partial^2 x} + K_y \frac{\partial^2 c}{\partial^2 y} +K_z \frac{\partial^2 c}{\partial^2 z} +S(r,t), 
    \label{final1}
    \tag{18a}
\end{align}
For the stress-driven VOC release, an impulse is considered to be the source, represented as:
\begin{align}
    S(r,t) = Q_0 \delta (x-x_0) \delta(y-y_0) \delta(z-z_0) \delta(t)
    \tag{18b}
    \label{source}
\end{align}
where ($x_0$, $y_0$, $z_0$) is the VOC releasing coordinate.
%This follows the initial condition as:
%\begin{align}
 %   c(r,0) = Q_0 \delta (x-x_0) \delta(y-y_0) \delta(z-z_0)
  %  \tag{2c}
   % \label{ini}
%\end{align}
Considering a moving frame, let us assume that $x=x-ut$, which modifies \eqref{final1} as:
\begin{align}
    \frac{\partial c}{\partial t}|_x = K_x \frac{\partial^2 c}{\partial^2 x} + K_y \frac{\partial^2 c}{\partial^2 y} +K_z \frac{\partial^2 c}{\partial^2 z} +S(r,t), 
    \label{final3}
    \tag{18c}
\end{align}
Imposing reflective boundary condition in $z$-direction and solving \eqref{final3} using Green's function \bluecite{yeh1975green}, we obtain:
\begin{align}
    c(\vec{r}, t) = \frac{Q_0}{(4 \pi t)^{3/2} \sqrt{K_x K_y K_z}} e^{-  \frac{(x-x_0)^2}{4K_x t}}  e^{- \frac{(y-y_0)^2}{4K_y t} } \nonumber \\
      \left[ e^{- \frac{{(z-z_0)}^2} {4K_z t}} + e^{- \frac{(z+z_0)^2}{4K_z t}}\right],
    \tag{18d}
    \label{green}
\end{align}
 Substituting $x=x-ut$ back in \eqref{green} and using the spread ($\sigma$) and eddy diffusivity ($K_i$) 
 relation, i.e., $\sigma^2 = 2 K_i t$ \bluecite{seinfeld2016atmospheric}, we obtain
\begin{align}
    c(\vec{r}, t) = \frac{Q_0}{(2 \pi)^\frac{3}{2} \sigma_x \sigma_y \sigma_z} e^{-\frac{1}{2}  \left( \frac{x-x_0-ut}{\sigma_x} \right)^2}  e^{-\frac{1}{2}\left( \frac{y-y_0}{\sigma_y} \right)^2} \nonumber \\
      \left[ e^{- \frac{{(z-z_0)}^2} {2\sigma_z^2}} + e^{- \frac{(z+z_0)^2}{2\sigma_z^2}}\right],
      \tag{19}
      \label{puff_ori}
\end{align}

\bibliography{references}
\bibliographystyle{IEEEtran}
\begin{comment}
\vspace{1pt}

\begin{IEEEbiography}[{\includegraphics[width=1in,height=1.25in,clip,keepaspectratio]{figures/bio_BM.jpg}}]{Bitop Maitra (Graduate Student Member, IEEE)}
received his B.Tech degree in Electronics and Communication Engineering from the University of Calcutta, Kolkata, India, in 2019 and the M.Tech in Biomedical Engineering from the National Institute of Technology, Rourkela, India, in 2021. He is pursuing his Ph.D. in Electrical and Electronics Engineering from Koç  University and is a research assistant at the Center for neXt-generation Communications (CXC). His research interests lie in abiogenesis, molecular communication, and related areas. 
\end{IEEEbiography}

\vspace{1pt}

\begin{IEEEbiography}[{\includegraphics[width=1in,height=1.25in,clip,keepaspectratio]{figures/bio_Ozgur_B._Akan.jpg}}]{Ozgur B. Akan (Fellow, IEEE)}
received his Ph.D degree from the Georgia Institute of Technology, Atlanta, GA, USA, in 2004. He is currently the Head of the Internet of Everything (IoE) Group with the Department of Engineering, University of Cambridge, UK, and the Director of the Centre for neXt-generation Communications (CXC), Koç University, Turkey. His research interests include wireless, nano, and molecular communications and the Internet of Everything. 
\end{IEEEbiography}
\end{comment}

\end{document}